\documentclass{aa}  

\usepackage{graphicx}
\usepackage[dvipsnames]{color}
\usepackage{array}
\usepackage{txfonts}
\newcommand{\mbar}{$M_\mathrm{bar}$}
\newcommand{\LCDM}{$\Lambda$CDM}
\newcommand{\kms}{$\mathrm{km}\,\mathrm{s}^{-1}$}
\newcommand{\aforty}{$\alpha$.40}
\newcommand{\gdtf}{$\alpha.$btfr}
\newcommand{\micron}{$\mu$m}
\newcommand{\kmsmpc}{$\mathrm{km}\,\mathrm{s}^{-1}\,\mathrm{Mpc}^{-1}$}

\newcommand{\fgas}{$f_\mathrm{gas}$}
\newcommand{\vrot}{$V_\mathrm{rot}$}
\newcommand{\vflat}{$V_\mathrm{flat}$}
\newcommand{\halpha}{$H_\alpha$}
\newcommand{\hbeta}{$H_\beta$}
\newcommand{\balmer}{$H_\alpha/H_\beta$}
\newcommand{\arcs}{$^{\prime\prime}$}
\newcommand{\Vhalo}{$V_\mathrm{h,max}$}

\begin{document}

\title{An accurate measurement of the baryonic Tully-Fisher relation with heavily gas-dominated ALFALFA galaxies}

\author{
E. Papastergis\inst{\ref{kapteyn}}\fnmsep\thanks{\textit{NOVA} postdoctoral fellow}, 
E.A.K. Adams\inst{\ref{astron}} 
\and J.M. van der Hulst\inst{\ref{kapteyn}}
}

\institute{
Kapteyn Astronomical Institute, University of Groningen, Landleven 12, Groningen NL-9747AD, The Netherlands \\ \email{papastergis@astro.rug.nl,vdhulst@astro.rug.nl}\label{kapteyn}
\and
ASTRON, the Netherlands Institute for Radio Astronomy, Postbus 2, Dwingeloo NL-7900AA, The Netherlands \\ \email{adams@astron.nl}\label{astron}
}

\titlerunning{BTFR of heavily gas-dominated galaxies}

\authorrunning{Papastergis et al.}


 
 \abstract{
 We use a sample of 97 galaxies selected from the Arecibo legacy fast ALFA (ALFALFA) 21cm survey to make an accurate measurement of the baryonic Tully-Fisher relation (BTFR). These galaxies are specifically selected to be heavily gas-dominated ($M_\mathrm{gas}/M_\ast \gtrsim 2.7$) and to be oriented edge-on. The former property ensures that the error on the galactic baryonic mass is small, despite the large systematic uncertainty involved in galactic stellar mass estimates. The latter property means that rotational velocities can be derived directly from the width of the 21cm emission line, without any need for inclination corrections.
 We measure a slope for the linewidth-based BTFR of $\alpha = 3.75 \pm 0.11$, a value that is somewhat steeper than (but in broad agreement with) previous literature results.
 The relation is remarkably tight, with almost all galaxies being located within a perpendicular distance of $\pm 0.1$ dex from the best fit line. The low observational error budget for our sample enables us to establish that, despite its tightness, the measured linewidth-based BTFR has some small (i.e., non-zero) intrinsic scatter. 
 We furthermore find a systematic difference in the BTFR of galaxies with ``double-horned'' 21cm line profiles --suggestive of flat outer galactic rotation curves-- and those with ``peaked'' profiles --suggestive of rising rotation curves. When we restrict our sample of galaxies to objects in the former category, we measure a slightly steeper slope of $\alpha = 4.13 \pm 0.15$.
 Overall, the high-accuracy measurement of the BTFR presented in this article is intended as a reliable observational benchmark against which to test theoretical expectations. 
 Here we consider a representative set of semi-analytic models and hydrodynamic simulations in the lambda cold dark matter (\LCDM) context, as well as modified Newtonian dynamics (MOND).
 In the near future, interferometric follow-up observations of several sample members will enable us to further refine the BTFR measurement, and make sharper comparisons with theoretical models.
}
  

   \maketitle
%

\section{Introduction}
\label{sec:intro}

The Tully-Fisher (TF) relation is one of the most fundamental scaling relations for disk galaxies. The relation owes its name to the seminal work of \citet{TullyFisher1977}, in which they report a fairly tight power-law relation between the rotational velocity of a galaxy and its optical luminosity. 
In that original work, the galactic rotational velocity was inferred from the spectral width of the 21cm emission line of atomic hydrogen (HI), while the galactic luminosity was measured from photographic plates.
Since then, the TF relation has become one of the most well-studied topics in extragalactic astronomy. Large programs of targeted HI observations and blind HI surveys have created samples of thousands of galaxies with kinematic information \citep[to name a few]{Geha2006,Springob2007,Meyer2008,Courtois2009,Haynes2011,Masters2014}. 
In addition, the availability of large multiwavelength photometric datasets has lead to measurements of the TF relation in almost every optical and near-infrared (NIR) band \citep[e.g.,][]{Karachentsev2002,Pizagno2007,Masters2008,Toribio2011,SaintongeSpekkens2011,Reyes2011,Hall2012,Sorce2013,Lagattuta2013,Neill2014}.
For smaller galactic samples, the TF relation has also been studied using measurements of their spatially resolved kinematics. Such datasets have become available thanks to targeted campaigns of interferometric observations in the HI line \citep[e.g.,][]{Sanders1996,VerheijenSancisi2001,McGaugh2005,deBlok2008,Begum2008a,Stark2009,Trachternach2009,Cannon2011,Hunter2012,Lelli2014,Ponomareva2016+}.

The TF relation has remained an extremely popular topic in the literature over many decades mainly for two reasons: First, it can be used to derive redshift-independent distance estimates for late-type galaxies. The rationale here is that the rotational velocity of a galaxy can be measured in a distance-independent way. From the measured rotational velocity, one can then infer the intrinsic luminosity of the galaxy, based on the TF relation measured for a ``reference'' sample \citep[e.g.,][]{Giovanelli1997a, Ferrarese2000,Freedman2011,Sorce2013,Neill2014}. A comparison of the inferred luminosity of the object to its measured brightness in the sky can then result in an estimate of its distance. 
When the TF relation is used as a distance indicator, the focus is on measurable quantities (such as luminosity in some specific band) and on samples of ``reliable'' objects. For example, many studies exclude dwarf galaxies from distance determinations, because the TF relation is often found to display increased scatter at the low-velocity end \citep[e.g.,][]{SaintongeSpekkens2011}. 
Systematic efforts to measure distances to nearby galaxies through the TF relation have lead to a variety of important scientific results, including measurements of the Hubble constant \citep[e.g.,][]{Giovanelli1997b,Masters2006} and maps of the peculiar velocity field and cosmic web in the local universe \citep[e.g.,][]{Masters2005,Courtois2013, CourtoisTully2015,Springob2016}.

Second, the TF relation can be used as a tool for studying galaxy formation \citep[e.g.,][]{vdBoschDalcanton2000}. This approach is based on the fact that the TF relation links a dynamical quantity that is related to the galactic dark matter content (the rotational velocity of a galaxy) to a quantity that is related to its baryonic content (such as luminosity). The TF relation can therefore be used as a constraint on any theoretical model that aims to link dark matter (DM) halos to observed galaxies. TF studies that aim to constrain galaxy formation models tend to have less strict selection criteria than distance studies,
because they benefit from probing a wide range in galactic properties. In addition, the focus of these studies is on quantities with a clear physical meaning, such as stellar or baryonic mass.
As a result, a lot of effort has recently been invested in the measurement of the stellar Tully-Fisher relation (STFR) and the baryonic Tully-Fisher relation (BTFR).  

The BTFR, in particular, has received a lot of attention \citep[to name a few]{Geha2006,Begum2008b,Stark2009,McGaugh2012,ForemanScott2012,Hall2012,Zaritsky2014,Lelli2016}, because it displays two puzzling characteristics: First, the BTFR is measured to follow a single power-law over a very wide range in galaxy mass \citep[e.g.,][]{McGaugh2000,Verheijen2001,Geha2006,McGaugh2012,ForemanScott2012,denHeijer2015,Lelli2016}. This behavior is surprising in the context of the \LCDM \ cosmological model. To reproduce other observed properties of the galactic population in \LCDM \ \citep[such as the ``baryonic mass function'' of galaxies;][]{Papastergis2012}, the baryon fraction of galaxies is expected to have a complex dependence on the mass of the host DM halo.
Second, the BTFR is measured to be surprisingly tight, to the point that some authors have argued for a zero intrinsic scatter relation \citep{Verheijen2001,McGaugh2012}. Once again, it is difficult to explain the relation's tightness theoretically, given that several sources of scatter for the relation are expected in \LCDM \ (scatter in halo concentrations and spins, scatter in galactic baryon fractions, etc.).
In response, a large number of theoretical works are devoted to modeling the BTFR as realistically as possible, and assessing how well the relation expected in \LCDM \ can reproduce the observed one \citep[to name a few]{Trujillo2011,Desmond2012,DesmondWechsler2015,SorceGuo2016,diCintioLelli2016,Sales2016}. Other works argue instead that the observed properties of the BTFR disfavor altogether a cosmological theory in which DM is the dominant mass component, and rather point to a modification of the law of gravity \citep[e.g., MOND;][]{Milgrom1983a}.

Clearly, the degree to which the BTFR can be used to constrain galaxy formation depends on how accurately the relation can be measured in the first place. The measurement of the BTFR is rather demanding observationally.
First, it requires measurements of the baryonic masses of galaxies, which are are typically computed from estimates of the the objects' stellar and atomic gas masses, $M_\mathrm{bar} = M_\ast + M_\mathrm{gas}$. The accuracy of baryonic mass measurements is usually limited by the uncertainty involved in stellar mass estimates derived from photometric data, which is typically 60\%-100\% \citep{Pforr2012}.
Second, it requires 21cm spectra for the objects of interest. Rotational velocities can then be derived by correcting the observed width of the spectral profile of each galaxy according to its inclination to the line-of-sight, $V_\mathrm{rot} = W/(2 \times \sin i)$. As a result, the accuracy of rotational velocity measurements is typically limited by systematic uncertainties in the measurement of galactic inclination.

In order to circumvent these observational limitations, we select here a sample of galaxies with two key properties: they are heavily gas-dominated ($ M_\mathrm{gas}/M_\ast \gtrsim 2.7$) and they are oriented edge-on.
The former characteristic ensures that the baryonic mass can be measured accurately, thanks to the fact that the atomic gas mass of a galaxy can be determined within an uncertainty of 10\%-15\%.
The latter property eliminates the need for inclination corrections when calculating the galactic rotational velocity. Compiling a modestly-sized sample of galaxies that satisfy these two conditions is very challenging, and was only made possible thanks to the large dataset of the Arecibo legacy fast ALFA survey \citep[ALFALFA survey;][]{Giovanelli2005}. This is because objects that have such a high gas-fraction (and also happen to be favorably oriented) are rare, in fact representing just about 4\% of the survey's detections.

The article is organized as follows: In Sec. \ref{sec:dataset} we describe in detail the selection process used to assemble the sample of gas-dominated and edge-on galaxies from the full ALFALFA catalog. In Sec. \ref{sec:derived_quantities} we describe our methods for estimating the physical quantities entering the BTFR (e.g., rotational velocity, stellar mass, gas mass) from the available observational data (e.g., optical photometry, 21cm spectra). We also describe the uncertainties associated with our estimates in the same section. In Sec. \ref{sec:btfr} we show our measurement of the BTFR, and elaborate on its characteristics. In Sec. \ref{sec:discussion} we compare our measurement of the BTFR with previous determinations in the literature, as well as with expectations from theory. In particular, we consider a number of semi-analytic models and hydrodynamic simulations in the \LCDM \ context, as well as the predictions of MOND. We end in Sec. \ref{sec:summary} by summarizing our work and briefly discussing our results.


\section{Dataset}
\label{sec:dataset}

\subsection{Sample selection}
\label{sec:sample_selection}

We use the publicly available ALFALFA dataset \citep[\aforty \ catalog;][]{Haynes2011} to select galaxies with suitable properties for an accurate measurement of the BTFR. 
In detail the selection criteria are the following:

\begin{enumerate}

\item \label{item:quality} \textit{High quality ALFALFA detections}: We select sources that are confidently detected by ALFALFA, $(S/N)_{HI} \geq 10$, and that are unambiguously classified as extragalactic objects (``code~1'' in \aforty). We also restrict ourselves to sources with estimated HI masses $M_{HI} > 10^7 \; M_\odot$. Lastly, we demand that the error on the measured velocity width (as reported in \aforty) is small, namely $W_{50,err}/W_{50} < 15$\%.

\item \label{item:oc} \textit{Sources with an optical counterpart in SDSS}: We only select sources cross-matched with a photometric object in the Sloan Digital Sky Survey (SDSS). This enables us to compute stellar masses for the ALFALFA galaxies. Note that many objects (but not all) have also a spectroscopic counterpart in SDSS.

\item \label{item:edge_on} \textit{Edge-on systems}: Galaxies are selected to be edge-on, so that no inclination corrections to the observed velocity widths are needed. In particular, edge-on galaxies are initially selected in an automated way, by demanding a low axial ratio in the SDSS $r$-band image ($b/a < 0.25$).

\end{enumerate}

\noindent
At this stage of the selection process we are left with 393 initial candidate galaxies. For these candidates we gather additional data in order to proceed with the next two steps of the selection process, including optical and NIR photometry from ancillary datasets (see \S\ref{sec:data_sources}).

\begin{enumerate}
\setcounter{enumi}{3}

\item \label{item:no_issues} \textit{Objects with no issues upon interactive inspection}: We visually inspect the SDSS image\footnotemark{} and the ALFALFA spectrum for each candidate galaxy, in order to weed out objects that are not suitable for the measurement of the BTFR.

\footnotetext{\texttt{skyserver.sdss.org/dr9/en/tools/chart/navi.asp}}

Based on the inspection of the SDSS images of candidate galaxies, we discard objects that were erroneously classified as edge-on based on their SDSS axial ratio. We also try to identify objects that are likely blended with other HI sources within the ALFALFA beam. Blends could contaminate the spectrum of the source of interest, and could therefore lead to inaccurate measurements of its HI flux and velocity width \citep{Jones2015}. Since the ALFALFA beam has a width of $\theta_\mathrm{fwhm} \approx 4^\prime$, we search the SDSS image of each candidate galaxy for possible contaminating sources within a radius of about $8^\prime$.
Moreover, we exclude galaxies that have disturbed optical morphologies, such as tidal features or strong asymmetries in their light profile. We then inspect the ALFALFA spectra of candidate galaxies to exclude objects with problematic HI profiles. These include cases where the spectrum is affected by radio frequency interference, or when the galactic lineprofile is highly asymmetric or has poorly defined edges. Note that, as long as the profile is well-behaved and reasonably symmetric, we do not further discriminate objects based on their profile shape (for example, ``double-horned'' vs. ``single-peaked'' profiles; refer to \S\ref{sec:profile_shape}). 

We assign to each candidate galaxy two subjective grades on a scale of 0-10, one based on the evaluation of its SDSS image and the other on the evaluation of its ALFALFA spectrum. Only galaxies with a total ``quality score'' of $\geq$17 out of 20 are included in the final sample (scores are assigned \textit{without} prior knowledge of each galaxy's position on the BTFR diagram).
Keep in mind that the overall quality assessment described above is subjective to some extent. For example, there can be cases where our target galaxy has companions that are nearby in the plane of the sky, but whose redshifts are not known (and therefore their status as a potential blend is uncertain). Figure \ref{fig:selection} shows five examples of quality assessment, and highlights the complications involved in the process. 

Even the decision to impose a quality score cutoff of 17/20 is largely subjective. Our quality score distribution is continuous, and thus drawing a line between the objects that are acceptable for the measurement of the BTFR and those that are not is arbitrary to some extent. We have chosen the specific cutoff value because we believe that it results in the best trade-off between the quality of the final sample and the total number of objects in it. In any case, we have verified that the results of this article do not depend sensitively on the adopted cutoff value.

\begin{figure*}[htb]
\centering
\includegraphics[scale=0.53]{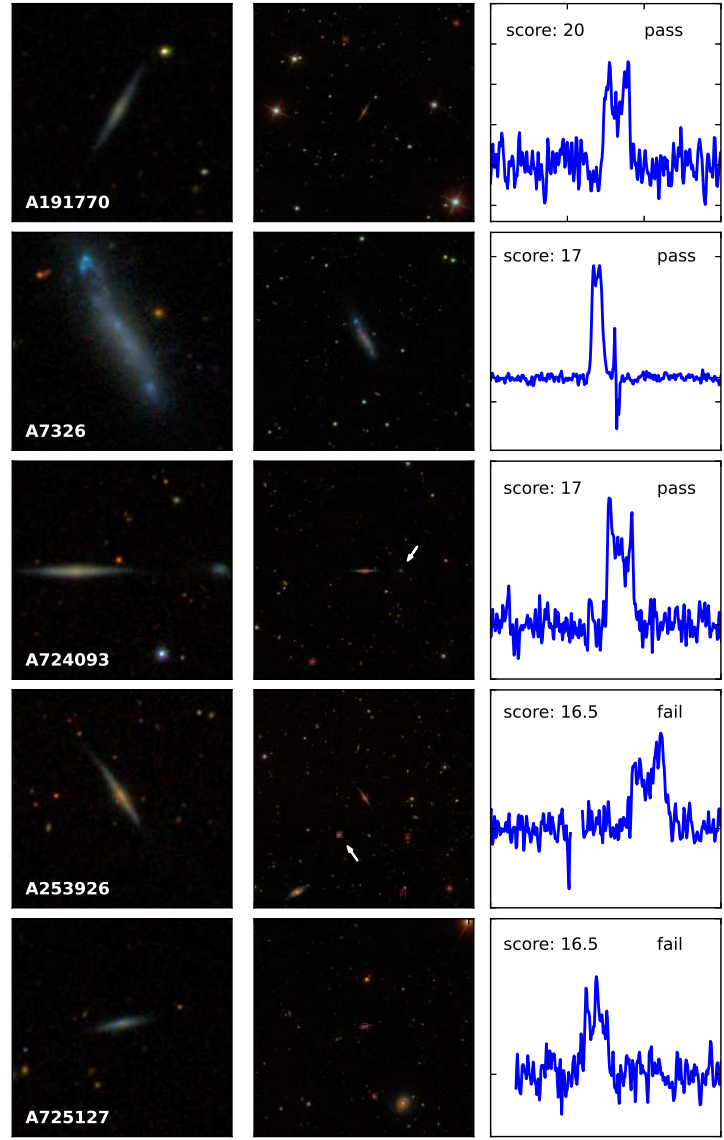}
\caption{ 
Quality inspection of candidate galaxies.
From left to right the three columns correspond to: (1) Close-up SDSS image of the candidate galaxy. In all cases the box has a size of $1.8^\prime \times 1.8^\prime$. North is up and East is to the left. 
(2) SDSS image of the broader area surrounding the candidate galaxy. Each box is $7.4^\prime \times 7.4^\prime$ in size, such that two ALFALFA beams could approximately fit side by side in the box. Red squares denote spectroscopic objects in the SDSS database, while arrows indicate potential blends. (3) ALFALFA spectrum of the HI line for each candidate galaxy. The $x$-axis corresponds to heliocentric velocity and has a fixed range of $1\,800$ \kms \ in all panels. The $y$-axis corresponds to flux density and its range varies from panel to panel. The annotated score refers to the subjective ``quality score'' described in item \ref{item:no_issues}, with a maximum value of 20. Galaxies with scores lower than 17 are eliminated from the final sample. Comments on individual galaxies (from top to bottom) follow below. \textit{AGC191770}: This galaxy has a perfect quality score. There are no companions around the galaxy and its HI line spectrum has no flaws. \textit{AGC7326}: This galaxy lies at a heliocentric velocity of $-167$ \kms, and its HI profile slightly overlaps with Galactic HI emission (visible as a subtraction artifact in the HI spectrum). However, the measurement of the HI width and flux for this specific galaxy was deemed reliable enough to be granted inclusion in the final sample. \textit{AGC724093}: This galaxy has a probable dwarf companion about $1^\prime$ to its West. Given that the companion is much dimmer than our target galaxy and that the physical association of the companion is not certain (the companion lacks an SDSS spectrum), the main target galaxy was nonetheless included in the final sample. \textit{AGC253926}: This galaxy has a spectroscopically confirmed companion that lies about $1^\prime$ to its Southeast, and is only slightly dimmer than the central galaxy. \textit{AGC725127}: The spectrum of this galaxy has a peculiar shape, with several peaks and a broad wing in the blueshifted edge of the profile. 
}
\label{fig:selection}
\end{figure*}

\item \label{item:gas_dominated} \textit{Gas-dominated galaxies}: Lastly, we select galaxies with a high ratio of HI to stellar mass, $M_{HI}/M_\ast > 2$. HI masses are obtained from ALFALFA fluxes (\S\ref{sec:atomic_gas}), while stellar masses are computed from a combination of several different methods (\S\ref{sec:stellar_mass}). After accounting for the cosmic abundance of helium and heavier elements, the gas fractions of these galaxies are $M_\mathrm{gas}/M_\ast \gtrsim 2.7$. Such high gas fractions\footnotemark{} ensure that an accurate determination of the galaxies' baryonic mass is feasible. 

\footnotetext{Keep in mind that the gas-richness selection criterion is somewhat fuzzy in practice: First, the stellar mass estimates carry significant uncertainty. As a result, a few galaxies that have true gas fractions near the cutoff value of $M_\mathrm{HI}/M_\ast = 2$ may cross the threshold (in either direction) simply due to measurement errors. Second, stellar masses are computed from a combination of several different methods (refer to \S\ref{sec:stellar_mass}). However, not all methods can be applied to every object, because many galaxies miss data that are necessary for one or another method (see \S\ref{sec:data_sources}).}

\end{enumerate}

Overall, the selection criteria \ref{item:quality}-\ref{item:gas_dominated} result in a final sample of 97 galaxies, which we refer to as the ``\gdtf'' sample. The \gdtf \ sample is the main focus of this article, and it is used in Sec. \ref{sec:btfr} to carry out a high-accuracy measurement of the BTFR.

\subsection{Data sources}
\label{sec:data_sources}

Atomic hydrogen properties and 21cm spectra are obtained from the \aforty \ catalog of the ALFALFA survey. The main data source for optical properties and images is instead the SDSS survey database. In particular, we make use of the databases of both the seventh data release \citep[DR7;][]{dr7_2009} and the ninth data release \citep[DR9;][]{dr9_2012} of SDSS. 

We also consider a number of ancillary optical and NIR datasets, which are used to obtain better stellar mass estimates for our galaxies. In particular, we use the NASA Sloan Atlas (NSA\footnotemark{}) catalog of galaxies. The NSA catalog is a comprehensive collection of data for galaxies in the nearby universe, mostly based on reprocessed SDSS photometric and spectroscopic data and on Galex ultraviolet data. The catalog contains several fields that are not available from SDSS, such as magnitudes for extended objects based on Sersic fits, emission line fluxes corrected for stellar absorption features, distance estimates based on a local flow model, and stellar mass estimates based on SED-fitting of the available photometric bands.

\footnotetext{\texttt{http://nsatlas.org}}

Moreover, we gather NIR photometry from the Two-Micron All Sky Survey (2MASS) and the Wide-field Infrared Survey Explorer (WISE) datasets. In particular, we have searched the 2MASS extended source catalog \citep[2MASS XSC;][]{Jarrett2000} for matches to all our initial 393 candidate galaxies. Objects with a crossmatch in 2MASS XSC have available photometry in the $J$, $H$ and $K_s$ bands (1.2, 1.7 and 2.2 \micron, respectively). Furthermore, we obtain WISE photometry for a subset of 120 out of the 393 candidate galaxies, which were determined to be gas-rich ($M_\mathrm{HI}/M_\ast \gtrsim 2$) according to preliminary estimates. Since WISE pipeline photometry is optimized for point sources, we do not use the magnitudes listed in the publicly available WISE catalog. Instead we have obtained dedicated resolved photometry measurements (T. Jarrett, private communication), based on the WISE images in the $W_1$ through $W_4$ bands (centered around 3.4, 4.6, 12 and 22 \micron \ respectively). Since we have not obtained WISE photometry for all the 393 initial candidate galaxies, the use of WISE data during step \ref{item:gas_dominated} of the selection process was not fully optimal.

While all \gdtf \ galaxies have an ALFALFA and an SDSS counterpart (selection criteria \ref{item:quality} \& \ref{item:oc}), they do not necessarily have data available in every ancillary dataset mentioned above. More specifically, 22 out of the 97 \gdtf \ galaxies lack an entry in the NSA catalog, while 26 lack WISE photometry of sufficient quality. Lastly, due to the limited depth of the 2MASS survey only 6 out of the 97 \gdtf \ galaxies have a cross-match in the XSC catalog.

\section{Derived quantities}
\label{sec:derived_quantities}

All incarnations of the TF relation refer to scaling laws among intrinsic galactic properties. The properties of interest are not directly observable (e.g., luminosity in various bands, stellar mass, rotational velocity), but have to be inferred from a set of measured quantities (e.g., apparent magnitude, redshift, observed velocity width). This process is often convoluted, and the uncertainties associated with these transformations typically dominate the error budget of the TF measurement. Below we describe how we derive the quantities that enter the measurement of the baryonic TF relation from the available observational data, and how we compute the associated uncertainties.

\subsection{Velocity width}
\label{sec:velocity_width}

Each \gdtf \ galaxy has an HI-line spectrum measured by the ALFALFA survey. From the spectrum we can measure the width of the lineprofile, which we denote by $W_{50}$. The subscript 50 refers to the fact that the width is measured between the two outermost points where the flux density of the profile falls to 50\% of the peak value. Since the galaxies used in this article have been specifically selected to be oriented edge-on, the measured value of velocity width, $W_{50}$ is the same as the intrinsic (i.e., deprojected) velocity width. We calculate the rest-frame velocity width, $W$, by correcting for the Doppler broadening of the profile:

\begin{eqnarray}
W = \frac{W_{50}}{(1+z_\odot)} \;\; . \;\;\;\;\;\; \mathrm{[edge-on~systems]}
\label{eqn:width}
\end{eqnarray}

\noindent
In the equation above, $z_\odot$ is the heliocentric redshift of the HI source. Note also that, unlike many literature studies of the BTFR, we do not subtract from $W$ the contribution of turbulent motions.

Despite the favorable orientation of the \gdtf \ galaxies, there are still some observational effects that can introduce a small uncertainty on the value of W. First, the measurement of $W_{50}$ is subject to statistical uncertainty, due to the finite signal-to-noise of the ALFALFA spectrum. The \aforty \ catalog reports a statistical error\footnotemark{} associated with the width measurement, which we denote by $\sigma_{W,\mathrm{stat}}$. The vast majority of our galaxies have statistical errors on $W_{50}$ of less than 8 \kms. Second, the inclination angle of a galaxy cannot be determined to be $i = 90^\circ$ with absolute certainty. This is because the galactic disk has some intrinsic thickness, the value of which is not generally known. 
However, due to the high inclination of the \gdtf \ galaxies, even a relatively large error in the inclination angle has minimal effects on the deprojection factor. For example, even if our galaxies had true inclination values of $i = 75^\circ$, the error on $W$ from assuming a perfectly edge-on orientation is only $\approx 3.5\%$.

\footnotetext{For some objects, the width error value reported in \aforty \ contains a systematic component related to the uncertainty in defining the edges of the spectral profile visually.}

In this article, the error on the intrinsic velocity width, $W$, of each galaxy is calculated as:

\begin{eqnarray}
\sigma_{W-}^2 & = & {\sigma_{W,\mathrm{stat}}}^2 \;\;\;\;\; \mathrm{and} \label{eqn:width_error_low} \\
\sigma_{W+}^2 & = & {\sigma_{W,\mathrm{stat}}}^2 + (0.035\cdot W)^2 \;\; .  \label{eqn:width_error_high}
\end{eqnarray}

\noindent
Note that the error term associated with the inclination uncertainty is only added to the positive errorbar. This is because any error in the inclination angle can never make the intrinsic width smaller than the value we calculate assuming a perfectly edge-on orientation (Eqn. \ref{eqn:width}).
The corresponding errors on the linewidth-derived velocity, $V_\mathrm{rot} = W/2$, are then simply: 

\begin{eqnarray}
\sigma_{V_\mathrm{rot}\pm} = \sigma_{W\pm}/2 \;\;\; .
\label{eqn:vel_error}
\end{eqnarray}


\subsection{Distance}
\label{sec:distance}

Distance estimates are necessary for computing distance-dependent intrinsic properties, such as the stellar and gaseous masses of galaxies. Our galaxies are all relatively nearby objects, but they nonetheless span a wide range in heliocentric velocity: from -167 \kms \ for UGC7326 to $\approx 12\,900$ \kms \ for AGC188761. This means that while for some of our galaxies simple Hubble distances may be appropriate, most of our objects require distance estimates based on a local flow model.   

In this article, we calculate distances and distance uncertainties based on the values provided by the \aforty \ and NSA catalogs. In particular, the adopted distance to each galaxy is the average of the distances listed in the two catalogs. Distance uncertainties are calculated based on the sum in quadrature of three error terms: the distance uncertainty reported in the \aforty \ catalog, the distance uncertainty reported in the NSA catalog, and the difference between the \aforty \ and NSA distance estimates. In particular, for each galaxy

\begin{eqnarray}
\label{eqn:dist}
D &=& \frac{D_{\alpha.40}+D_\mathrm{NSA}}{2}  \;\;\;\;\;  \mathrm{and} \\
\sigma_D^2 &=& \sigma_{D,\alpha.40}^2 + \sigma_{D,\mathrm{NSA}}^2 + \left(\frac{D_{\alpha.40}-D_\mathrm{NSA}}{2} \right)^2 \;\;  . 
\label{eqn:dist_err}
\end{eqnarray}   

\noindent
Let us now briefly comment on each of the terms entering Eqns. \ref{eqn:dist} \& \ref{eqn:dist_err}.  

The \aforty \ catalog assigns distances to relatively distant galaxies ($V_\odot \gtrsim 6\,000$ \kms) based on Hubble's law. In particular, distances are assigned based on a galaxy's recessional velocity in the CMB frame, $D_{\alpha.40} = V_\mathrm{CMB}/H_0$, assuming a Hubble constant of $H_0 = 70$ \kmsmpc. The \aforty \ catalog does not report errors associated with Hubble distances, so we adopt here the error corresponding to a Hubble constant uncertainty of $\sigma_{H_0} = 3$~\kmsmpc \ \citep[e.g.,][]{Riess2011,Freedman2012,Planck2014}. We then add in quadrature a second error term to the Hubble distances, reflecting a peculiar velocity of $V_\mathrm{pec} = 160$ \kms \ \citep{Masters2005}. 

For more nearby galaxies ($V_\odot \lesssim 6\,000$ \kms), the \aforty \ catalog assigns distances based on the flow model of \citet{Masters2005}. In these cases we compute a distance error by adding in quadrature the error reported in \aforty \ (which is associated with the uncertainty in the flow model) and the Hubble constant error. Lastly, the \aforty \ catalog makes manual assignments of galaxies to known groups (including Virgo and its substructures). In these cases we assume no peculiar velocity errors and only consider errors from the uncertainty in $H_0$.

A similar distance assignment scheme is also followed by the NSA catalog. In particular,  relatively distant galaxies ($V_\odot \gtrsim 6\,000$ \kms) are assigned Hubble distances, based on their recessional velocity in the Local Group frame of reference, $D_\mathrm{NSA} = V_\mathrm{LG}/H_0$. More nearby objects are assigned distances based on the flow model of \citet{Willick1997}. The NSA catalog lists a distance uncertainty for all galaxies, which we use in Eqn. \ref{eqn:dist_err} as reported. 
As far as the third term in Eqn. \ref{eqn:dist_err} is concerned, differences in distance estimates for farther galaxies stem from the fact that the two catalogs use two different velocity reference frames (CMB vs. Local Group). As a result, the distance offset between the two catalogs is a strong function of direction on the sky. For nearby galaxies, differences reflect instead the fact that the two catalogs use two different flow models.
Note that the NSA catalog does not make galaxy assignments to groups. If a galaxy is assigned to a group by \aforty, we ignore the NSA distance estimate and fix the second and third terms in Eqn. \ref{eqn:dist_err} to typical values. We follow the same procedure also for galaxies that are lacking an entry in the NSA catalog.

Overall, the fractional distance uncertainty for the majority of \gdtf \ galaxies is low, $\sigma_D/D < 10\%$. This is not always the case however for low-mass galaxies, which are typically located nearby. For these objects peculiar motions can lead to quite a large distance uncertainty. Among the \gdtf \ galaxies, UGC8575 (located at an estimated distance of 15.1 Mpc) suffers from the largest fractional distance uncertainty of $\sigma_D/D \approx 35\%$.

\subsection{Atomic gas mass}
\label{sec:atomic_gas}

We calculate HI masses from the HI-line flux measured by ALFALFA, $S_{HI}$. In particular, we use the relation:

\begin{eqnarray}
M_{HI}(M_\odot) = 2.356\times10^5 \cdot S_{HI}(\mathrm{Jy}\:\mathrm{km}\,\mathrm{s}^{-1}) \cdot D^2(\mathrm{Mpc}) \;\; ,
\label{eqn:hi_mass}
\end{eqnarray} 

\noindent 
where $D$ is the distance to the galaxy (calculated as described in \S\ref{sec:distance}). We assume two sources of error for the measured HI flux: The first is statistical in nature, and is due to the finite signal-to-noise of the ALFALFA spectrum. We adopt as our statistical flux error the value listed in the \aforty \ catalog. The second is systematic and relates to the accuracy of the absolute calibration of the ALFALFA fluxes. We adopt a fixed value of 10\% for this systematic error component (M.P. Haynes, private communication).

We calculate neutral atomic gas masses as 

\begin{eqnarray}
M_\mathrm{gas} = 1.33 \times M_{HI} \;\;\; . 
\end{eqnarray}

\noindent
The $1.33$ factor is used to account for the primordial abundance of helium and heavier elements. Accounting for helium in this way is standard practice in BTFR studies \citep[e.g.,][]{McGaugh2012}. Lastly, we would like to note that we do not attempt to correct for possible loss of 21cm flux due to HI self-absorption, as the magnitude of the effect is still debated in the literature.

\subsection{Stellar mass}
\label{sec:stellar_mass}

The computation of galactic stellar mass can be divided into two subprocesses: first, the determination of the galactic luminosity within one (or more) photometric filters; second, the calculation of the mass-to-light ratio (i.e., the conversion from luminosity to stellar mass). 

The former computation is relatively straightforward. For any given photometric band $x$, we use the apparent magnitude, $m_x$, and the distance, $D$, to calculate the absolute magnitude

\begin{eqnarray}
M_x &=& m_x - 25 -5 \cdot \log_{10}D(\mathrm{Mpc}) \;\; .
\label{eqn:absolute_mag}
\end{eqnarray}

\noindent 
Then we use the absolute magnitude of the Sun in the same band, $M_{x,\odot}$, to compute the luminosity in terms of solar luminosities

\begin{eqnarray}
L_x(L_\odot) &=&  10^{(M_x -M_{\odot,x})/2.5}    \;\; .
\label{eqn:luminosity}
\end{eqnarray}  

\noindent
The process above is directly analogous to Eqn. \ref{eqn:hi_mass}, but its similarity is concealed by the fact that optical fluxes are reported in terms of magnitudes rather than in linear flux units. The magnitudes used in Eqns. \ref{eqn:absolute_mag} \& \ref{eqn:luminosity} should be corrected for the Doppler shift of the stellar spectrum ($k$-correction), and for foreground Milky Way extinction \citep{Schlegel1998}. Unless otherwise specified, magnitudes in this article always refer to $k$-corrected and MW extinction-corrected values.  

Apparent magnitudes, $m_x$, can be measured fairly accurately. For example, typical statistical and calibration uncertainties for point-source magnitudes in the $g,r,i$ bands of SDSS are $< 2$\% \citep{Padmanabhan2008}. The largest uncertainty affecting the photometry of galaxies actually comes from differences in the operational definition of apparent magnitude for extended sources. For example, the SDSS catalog reports two types of magnitudes for extended objects that differ in their measurement methodology: \texttt{petroMag} and \texttt{modelMag}\footnotemark{}. The two different methods can have typical offsets of $\lesssim 15\%$ in terms of linear flux for the same galaxy. The accuracy of extended source photometry also depends on how well foreground and background contaminating sources are deblended. 

\footnotetext{For details on the calculation of SDSS Petrosian magnitudes (\texttt{petroMag}) and model magnitudes (\texttt{modelMag}) please refer to \texttt{http://www.sdss3.org/dr8/algorithms/magnitudes.php}.}

Occasionally, incorrect deblending can affect the target source itself, by ``shredding'' it into multiple photometric objects. Shredding mostly affects galaxies that are extended on the sky or that have patchy morphologies, and can lead to a large underestimate of the total optical flux. We try to identify cases where shredding could be an issue during the visual inspection step of the selection process (item \ref{item:no_issues}). In cases where shredding is suspected to be an important issue, we perform manual aperture photometry using the \texttt{Aladin} visualization tool\footnotemark{}. 
A second source of error in the computation of intrinsic luminosities (Eqns. \ref{eqn:absolute_mag} \& \ref{eqn:luminosity}) is the uncertainty in distance estimates. This term can sometimes be dominant,  especially in the case of low-mass galaxies which are typically located nearby (see \S\ref{sec:distance}).

\footnotetext{\texttt{aladin.u-strasbourg.fr}}

On the other hand, the conversion from luminosity to stellar mass is definitely a much more involved and uncertain process. The main idea is that the ratio of fluxes in different photometric bands (i.e., the observed galactic colors) contain information about the properties of the galactic stellar population (age, metallicity, dust extinction, etc.). One can therefore create a set of stellar population models covering a wide range of physical properties, and find the models that best reproduce the observed colors. The mass-to-light ratio of the best fitting stellar population can then be used to convert the observed galactic luminosity into stellar mass. In its most general form, this process is called spectral energy distribution fitting (SED-fitting). Note that even though the process above seems straightforward, several strong degeneracies exist among model parameters (e.g., between star formation rate and extinction or between star formation rate and metallicity), that complicate the inference of the true stellar population properties from photometry.

Some simplified approaches are also possible: For example, one can use a single color (e.g., $g-i$) as a proxy for the mass-to-light ratio derived from full SED-fitting of several photometric bands. There exist many calibrations of such single-color prescriptions in the literature \citep[e.g.,][to name a few]{BelldeJong2001,Bell2003,Taylor2011}. Due to their ease of use, single-color prescriptions have been widely popular in the literature. Alternatively, one can seek a photometric band where the mass-to-light ratio is fairly constant, regardless of the detailed properties of the stellar population. This approach is usually applied to photometry at near-infrared (NIR) wavelengths, in which case stellar mass is taken to be roughly proportional to the NIR luminosity (\citealp[e.g.,][]{Bell2003,McGaugh2015, Martinsson2013,Angus2016}; but see also \citealp{Eskew2012,Cluver2014}). Photometry at NIR wavelengths has also the added advantage of being less affected by dust extinction compared to optical photometry. 

Regardless of the details, the conversion between luminosity and stellar mass is a process that is subject to large systematic uncertainties. This statement is particularly true for actively star-forming galaxies, such as those in the \gdtf \ sample. An additional complication for the \gdtf \ galaxies is the fact that they are oriented edge-on, and thus internal extinction may represent a source of bias when computing stellar masses. We address this concern in Appendix \ref{sec:appendix_extinction}, where we show that \gdtf \ galaxies do not display higher levels of internal extinction than any typical galactic sample.  

Quantifying the error associated with stellar mass estimates is itself quite challenging, and different error values are quoted by different authors \citep[e.g.,][]{GallazziBell2009,Conroy2009, ConroyGunn2010,Maraston2010,Pforr2012,Torrey2015}. When a comprehensive set of methods is considered, a realistic value for the uncertainty on the stellar mass of an individual object is $\sim$ 0.2-0.3 dex, while systematic biases between different methods can also have a similar magnitude \citep{Pforr2012}. An important point to keep in mind is that stellar mass estimates depend not only on the type of observational data used to derive them (e.g., part of spectrum, type and number of photometric filters, etc.), but also on the theoretical framework used to model the stellar population. In most cases, the error on the stellar mass estimate is dominated by theoretical uncertainties, and it cannot therefore be reduced by improvements in the photometry.

In view of the considerations above, we refrain from adopting a particular stellar mass method ``of choice'' in this article. We consider instead a variety of different stellar mass methods, producing several estimates of the stellar mass for each galaxy.
This approach has two advantages: First, it permits us to identify cases where one of the methods fails, either due to data issues or due to to the method's limitations. Second, it enables us to use the scatter between the individual stellar estimates as a gauge of the overall stellar mass uncertainty, on an object-by-object basis.

More specifically, the methods that we consider in this article are the following:

\begin{enumerate}[i.]

\item \label{item:sed} We calculate stellar masses based on SED-fitting of the five SDSS bands, according to the method outlined in \citet[\S4.1-4.2]{Huang2012a}. In particular, the methodology is similar to the one used by \citet{Salim2007}, but can also account for bursts in the star-formation history of the galaxy. Such bursts are expected to be ubiquitous in gas-rich galaxies \citep[e.g.,][]{Lee2009}. The method adopts a \citet{Chabrier2003} initial mass function (IMF), and is fitted on SDSS \texttt{model} magnitudes. We denote this mass by $M_\ast^\mathrm{(SDSS)}$.

\item \label{item:taylor} We calculate stellar masses from $i$-band luminosities, with stellar mass-to-light ratios inferred from the galactic $g-i$ color \citep[Eqn. 7]{Taylor2011}. Once again, we use this method in conjunction with SDSS \texttt{model} magnitudes. We denote this mass by $M_\ast^{(g-i)}$.

\item \label{item:nsa} We use the stellar masses reported by the NSA catalog. These stellar masses are also based on SED-fitting, in particular of the five SDSS optical bands ($u,g,r,i,z$) plus the two Galex ultraviolet bands ($NUV,FUV$) when available. Note that the NSA fitting process does not use the magnitudes listed in the SDSS catalog, but instead uses Sersic magnitudes\footnotemark{}. In addition, the NSA stellar masses are based on stellar population models that are different from the models used in method \ref{item:sed}. We denote this mass by $M_\ast^\mathrm{(NSA)}$.  

\footnotetext{Sersic magnitudes are derived from fits of two-dimensional, one-component Sersic models to the SDSS image. The Sersic index, half-light radius, position angle, axial ratio and flux are free parameters which are determined by the fit.}

\item \label{item:wise} We compute stellar masses from NIR photometry provided by WISE. In particular, we calculate galactic luminosities in the $W_1$ band, adopting a constant mass-to-light ratio of $M/L_{W1} = 0.45 \; M_\odot/L_\odot$ \citep{McGaugh2015}. We denote this mass by $M_\ast^\mathrm{(WISE)}$.

\item \label{item:2mass} We derive stellar masses from $K_s$ band photometry provided by 2MASS. In particular, we use the extrapolated total magnitudes reported in the XSC catalog, and we adopt a constant mass-to-light ratio of $M/L_K = 0.6 \; M_\odot/L_\odot$ \citep{Bell2003,McGaugh2015}. We denote this mass by $M_\ast^\mathrm{(2MASS)}$.


\end{enumerate}

We then derive a ``consensus'' stellar mass value for each object, as the logarithmic average of the values calculated through the set of methods \ref{item:sed}-\ref{item:2mass}:

\begin{eqnarray}
\log M_\ast = <\log M_\ast^\mathrm{(i-v)}>   \;\;\; .
\end{eqnarray}

\noindent

We derive an error on the mass-to-light ratio of each galaxy, by adding in quadrature the scatter in the method estimates \ref{item:sed}-\ref{item:2mass} and a fixed systematic error term:

\begin{eqnarray}
{\sigma_{M_\ast/L}}^2 = {\sigma_{\log M_\ast}^\mathrm{(i-v)}}^2  +  {\sigma_\mathrm{sys}}^2 \;\;\; .
\label{eqn:m_to_l_error}
\end{eqnarray}    

\noindent
We adopt the value of $\sigma_\mathrm{sys} = 0.25$ dex, in order to capture the typical offsets and scatter in the recovery of stellar masses from SED-fitting found by \citet{Pforr2012}. In cases where there are only one or two available stellar mass estimates, the term $\sigma_{\log M_\ast}^\mathrm{(i-v)}$ in Eqn. \ref{eqn:m_to_l_error} 
can be artificially low (e.g., zero for the case of one estimate). In these cases, we manually set this term 
to the typical value of 0.11 dex.






\subsection{Baryonic mass}
\label{sec:baryonic_mass}

We calculate baryonic masses, \mbar, as the sum of the galactic stellar and atomic gas mass (refer to \S\ref{sec:atomic_gas} \& \ref{sec:stellar_mass}): 

\begin{eqnarray}
M_\mathrm{bar} = M_\ast + M_\mathrm{gas}  \label{eqn:mbar} \;\;\; .
\label{eqn:mbar_def}
\end{eqnarray}

\noindent
If we define the gas fraction as \fgas$\equiv M_\mathrm{gas}/M_\ast$, then the baryonic mass can be also expressed as 

\begin{eqnarray}
M_\mathrm{bar} = M_\ast \times (1+f_\mathrm{gas}) = M_\mathrm{gas} \times (1+f_\mathrm{gas}^{-1})  \label{eqn:mbar_alter} \;\;\; .
\label{eqn:mbar_altern}
\end{eqnarray}

\noindent 
Given that \gdtf \ galaxies have gas fractions \fgas$\gtrsim 2.7$, Eqn. \ref{eqn:mbar_altern} implies that their baryonic mass is equal to their atomic gas mass up to a maximum correction of $\approx$35\%.

Note that \mbar \ as defined above does not actually account for the total baryonic mass present within the extent of the galaxy's virial radius. For example, baryonic components such as ionized gas and molecular gas are not included in Eqn. \ref{eqn:mbar_def}. In this article, the term ``baryonic mass'' refers to the sum of two major and easily observable components, namely stars and neutral atomic gas. Nonetheless, these two components still account for the vast majority of baryonic material associated with the galactic disk.

\subsection{Baryonic mass error budget}
\label{sec:baryonic_error}

The error on the logarithmic baryonic mass is given by  

\begin{eqnarray}
\sigma_{\log_{10}(M_\mathrm{bar})} = \frac{1}{\ln(10)} \; \frac{\sigma_{M_\mathrm{bar}}}{M_\mathrm{bar}} \;\; .
\label{eqn:mbar_err_1}
\end{eqnarray}

\noindent
Using Eqns. \ref{eqn:mbar} \& \ref{eqn:mbar_alter} we can re-write the fractional error on the baryonic mass as:

\begin{eqnarray}
\left( \frac{\sigma_{M_\mathrm{bar}}}{M_\mathrm{bar}} \right)^2 & = & \left( \frac{\sigma_{M_\ast}}{M_\mathrm{bar}} \right)^2 + \left( \frac{\sigma_{M_\mathrm{gas}}}{M_\mathrm{bar}} \right)^2  \notag \\
& = & \left( \frac{1}{1+f_{gas}}\frac{\sigma_{M_\ast}}{M_\ast} \right)^2 + \left( \frac{1}{1+f_\mathrm{gas}^{-1}}\frac{\sigma_{M_\mathrm{gas}}}{M_\mathrm{gas}} \right)^2   
\label{eqn:mbar_err_2} \;\; .
\end{eqnarray}

\noindent
The equation above explicitly demonstrates why gas-dominated galaxies have low fractional errors on their measured baryonic mass. Typically, the largest source of uncertainty in the computation of \mbar \ stems from the uncertainty in the galactic stellar mass. However, in Eqn. \ref{eqn:mbar_err_2} the stellar mass error term is suppressed by a factor of $1+f_\mathrm{gas}$. For the gas-dominated galaxies in the \gdtf \ sample, the suppression factor is $\gtrsim$3.7.

Let us now elaborate on each error term in Eqn. \ref{eqn:mbar_err_2}. First, the fractional error on the stellar mass can be calculated based on Eqns. \ref{eqn:absolute_mag} \& \ref{eqn:luminosity} as:

\begin{eqnarray}
\left( \frac{\sigma_{M_\ast}}{M_\ast} \right)^2 = \left( 10^{\sigma_m/2.5} \right)^2 + \left( 2\frac{\sigma_D}{D} \right)^2 + \left( \frac{\sigma_{M_\ast/L}}{M_\ast/L} \right)^2  
\label{eqn:mstar_err} \;\; .
\end{eqnarray}    

\noindent
In the equation above\footnotemark{}, $\sigma_m$ is the error in the apparent magnitude, for which we adopt a representative value of 0.15 mag (equivalent to an uncertainty on the linear flux of $\approx$15\%); $\sigma_D$ is the uncertainty in the distance, which we calculate as described in \S\ref{sec:distance}; $\sigma_{M_\ast/L}$ is the uncertainty in the stellar mass-to-light ratio, which we calculate according to Eqn. \ref{eqn:m_to_l_error} (refer to \S\ref{sec:stellar_mass}).

\footnotetext{Strictly speaking, Eqn. \ref{eqn:mstar_err} is valid only in cases where the stellar mass is calculated based on the galactic luminosity in some specific band, multiplied by a stellar mass-to-light ratio in the same band (i.e., only for methods \ref{item:taylor}, \ref{item:wise}  \& \ref{item:2mass}). On the other hand, methods \ref{item:sed} \& \ref{item:nsa} use multi-band photometric fitting to estimate stellar masses, and therefore the total error on the stellar mass cannot be neatly separated into the individual terms of Eqn. \ref{eqn:mstar_err}. However, Eqn. \ref{eqn:mstar_err} should be roughly applicable to these methods as well, because the uncertainty in the mass-to-light ratio is dominant compared to the photometric errors.}

Second, the fractional error on the gas mass in Eqn. \ref{eqn:mbar_err_2} can be evaluated based on Eqn. \ref{eqn:hi_mass} as:   

\begin{eqnarray}
\left( \frac{\sigma_{M_\mathrm{gas}}}{M_\mathrm{gas}} \right)^2 = \left( \frac{\sigma_{S_{HI}}}{S_{HI}} \right)^2 + \left( 2\frac{\sigma_D}{D} \right)^2  \label{eqn:mhi_err} \;\;\;  .
\end{eqnarray}

\noindent
Here, $\sigma_{S_{HI}}/S_{HI}$ is the fractional error in the measured HI flux, which we calculate as described in \S\ref{sec:atomic_gas}.

Overall, the median error on the baryonic mass of \gdtf \ galaxies is remarkably small, $ \widetilde{\sigma}_{\log(M_\mathrm{bar})} = 0.09$ dex. However, baryonic mass errors can be quite a bit larger for some individual objects, especially those that have a large uncertainty on their distance. UGC8575 is the \gdtf \ galaxy with the largest baryonic mass error of 0.31 dex.

\section{The baryonic Tully-Fisher relation of gas-dominated galaxies}
\label{sec:btfr}

\begin{figure}[htb]
\centering
\includegraphics[width=\columnwidth]{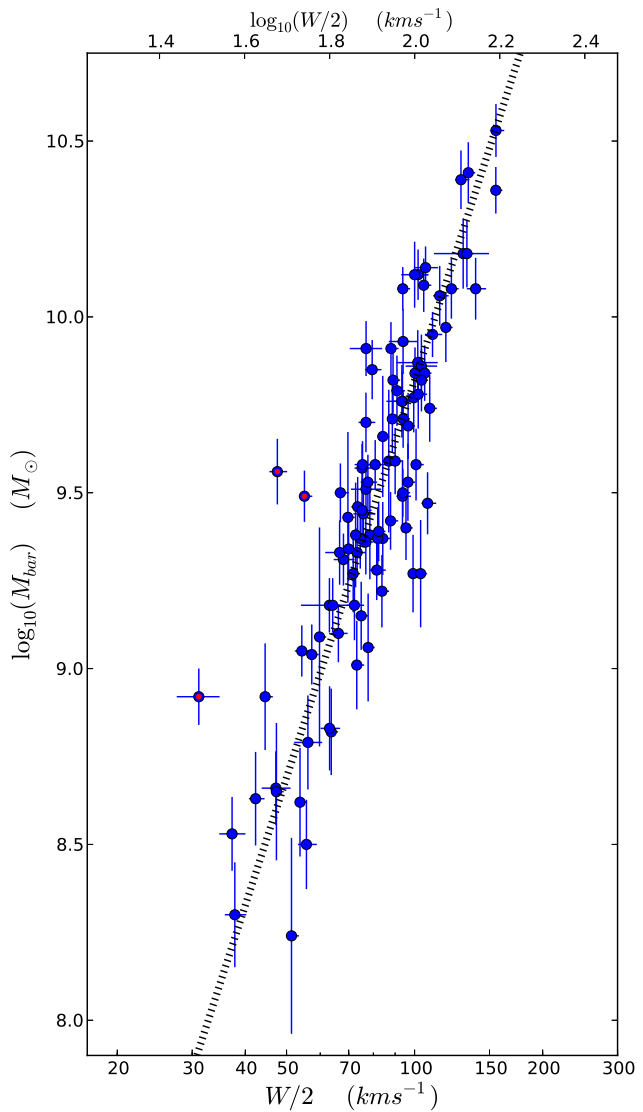}
\caption{ 
BTFR of gas-dominated, edge-on galaxies.
Blue circles represent the 97 galaxies in the \gdtf \ sample, which is selected from the ALFALFA survey (see \S\ref{sec:sample_selection}). The values of rotational velocity and baryonic mass, as well as their errors, have been computed as described in Sec. \ref{sec:derived_quantities}. The three outliers to the relation are marked by a red dot superimposed on their symbol. The thick dotted line represents the best linear fit to the BTFR (excluding outliers), which has a slope of $\alpha = 3.58 \pm 0.11$.  
}
\label{fig:btfr}
\end{figure}

Figure \ref{fig:btfr} shows the baryonic Tully-Fisher relation measured with the \gdtf \ sample of gas-dominated and edge-on galaxies. The x-axis corresponds to rotational velocity, as measured by the width of the HI lineprofile, \vrot$= W/2$ (see \S\ref{sec:velocity_width}). The y-axis corresponds to baryonic mass, defined here as the sum of stellar and atomic gass mass, \mbar$= M_\ast + M_\mathrm{gas}$ (see \S\ref{sec:baryonic_mass}). Given the properties of the \gdtf \ sample, there is very little room for observational measurement errors and measurement systematics to affect the presented BTFR. As a result, our measurement can be used as an observationally robust benchmark for the linewidth-based BTFR, against which to compare theoretical models and simulations.

\begin{figure*}[htb]
\centering
\includegraphics[scale=0.52]{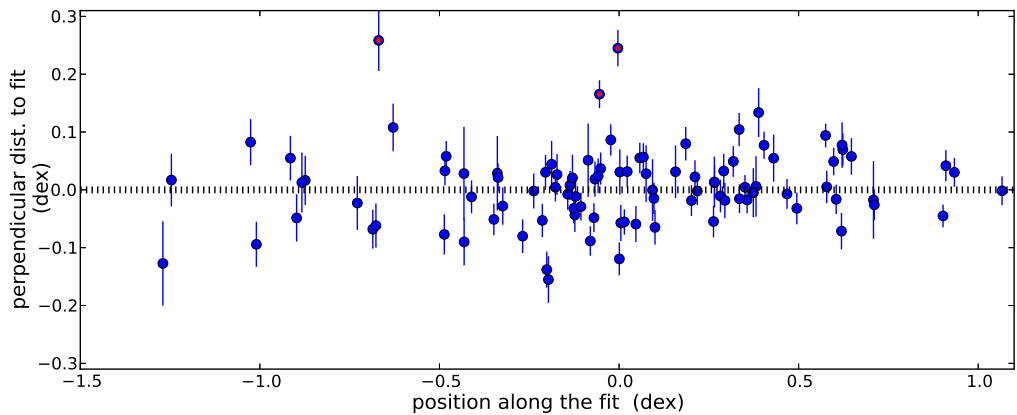}
\caption{ 
Positions of \gdtf\ galaxies with respect to the best fit line to the BTFR.
The blue datapoints with errorbars represent the \gdtf \ galaxies, with outliers marked with a red dot (same as Fig. \ref{fig:btfr}).  The $x$- and $y$-coordinate of each datapoint corresponds respectively to its position along the best fit line to the BTFR and perpendicular to it (refer to Fig. \ref{fig:btfr}). The errorbars are derived by projecting the errors on both the velocity and the baryonic mass onto the direction perpendicular to the fit line.
}
\label{fig:perp+parallel}
\end{figure*}

Our sample probes the BTFR over the range of intermediate masses, $M_\mathrm{bar} \approx 10^{8.5} - 10^{10.5} \; M_\odot$, and corresponding velocities in the range  $V_\mathrm{rot} \approx 40 - 150$ \kms. The BTFR shown in Fig. \ref{fig:btfr} displays a well-defined ``main body'', with just three ``low-side outliers''. We perform a maximum likelihood orthogonal fit to the BTFR presented in Fig. \ref{fig:btfr}, excluding the three outliers. We obtain a best fit slope of $\alpha = 3.75 \pm 0.11$. This value is somewhat steeper than (but in general agreement with) the values obtained by previous studies for the linewidth-based BTFR (e.g., \citealp{Hall2012,Zaritsky2014,McGaugh2012}; refer to \S\ref{sec:btfr_vs_others}). In Appendix \ref{sec:fit_details} we describe in detail our fiducial fitting method, which was used to derive the best fit parameters reported in this article. Keep in mind that the parameters of the best fit line to the BTFR do not only depend on the observational datapoints, but also on the fitting method employed \citep{Bradford2016}. For this reason, we show in Appendix \ref{sec:alternative_fits} results obtained with different fitting methodologies.



\begin{figure}[htb]
\centering
\includegraphics[scale=0.47]{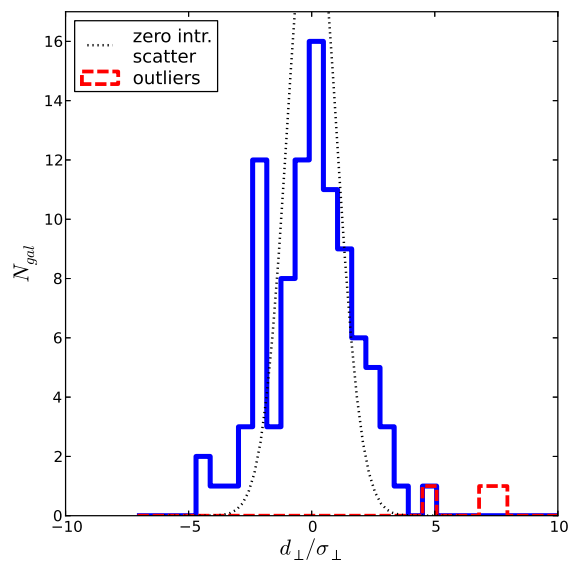}
\caption{ 
Histogram of normalized perpendicular distances for \gdtf \ galaxies.
The blue histogram represents the distribution of $d_{\perp,i}/\sigma_{\perp,i}$, which is the perpendicular distance of a \gdtf \ datapoint from the fit line, normalized by the perpendicular error for the specific datapoint (refer to Fig. \ref{fig:perp+parallel}). 
The thin dotted line is a standard normal distribution, which is the expected distribution for a zero intrinsic scatter relation.
}
\label{fig:histogram}
\end{figure}

Moreover, the relation is remarkably tight in the direction perpendicular to the fit line. This is illustrated by Figure \ref{fig:perp+parallel}, which shows that most galaxies are located within a distance of $\pm0.1$ dex from the fit line in the perpendicular direction. The standard deviation of the perpendicular distances of \gdtf \ datapoints from the fit line is only $\sigma_\perp = 0.056$ dex; even if we ignore observational errors altogether, this measured value sets an upper limit on the \textit{intrinsic} scatter of the BTFR in the perpendicular direction, namely $\sigma_{\perp,\mathrm{intr}} < \sigma_\perp = 0.056$~dex. Figure \ref{fig:perp+parallel} can also be used to further assess whether the observed scatter in the datapoints reflects  solely the observational errors (or, in other words, if the datapoints are consistent with a zero intrinsic scatter relation). In particular, one can consider the distance of each datapoint from the fit line in the perpendicular direction, $d_{\perp,i}$, and normalize it by the perpendicular error for that datapoint, $\sigma_{\perp,i}$. The perpendicular error is calculated from the sum in quadrature of the mass and velocity errors, after each of them has been projected on the direction perpendicular to the fit line (see Appendix \ref{sec:fit_details}). If the observed total scatter were entirely due to measurement errors, then the histogram of $d_{\perp,i}/\sigma_{\perp,i}$ would follow a standard normal distribution. 
Figure \ref{fig:histogram} presents the result of this test for the full \gdtf \ sample. According to the figure, the spread in values of $d_{\perp,i}/\sigma_{\perp,i}$ is larger than expected from a standard normal distribution (although not by much). Figure \ref{fig:histogram} points therefore to a linewidth-based BTFR with a small, but non-zero, value of intrinsic scatter. The fact that the BTFR of the \gdtf \ sample has non-zero intrinsic scatter is further confirmed by the analysis performed in Appendix \ref{sec:alternative_fits}. In particular, when intrinsic scatter is included in the linear model as a free parameter, the $\sigma_{\perp,\mathrm{intr}} = 0$ case is excluded at high significance (see Fig. \ref{fig:chi2_ellipse_scatter}).

An important caveat regarding the result of Figure \ref{fig:histogram} is that, given the tightness of the relation, the outcome relies critically on having accurate estimates for the observational errors. The properties of the \gdtf \ galaxies help keep low not only the observational errors, but also the uncertainty with which the observational errors themselves are determined. Nonetheless, there are still physical quantities entering the measurement of the BTFR for which the associated observational uncertainties are difficult to quantify precisely (most notably distances and stellar masses, refer to \S\ref{sec:distance} \& \S\ref{sec:stellar_mass}). In general, any underestimate of the observational errors leads to an overestimate of intrinsic scatter, and vice versa.

\subsection{Dependence of the BTFR on the shape of the HI line profile}
\label{sec:profile_shape}

Based on theoretical arguments, the BTFR is expected to reflect a relation between the baryonic mass of a galaxy, \mbar, and the mass of its host halo, $M_h$. In general, $M_h$ is not observable, but cosmological simulations have shown that it is strongly correlated with the peak circular velocity of the halo's rotation curve, \Vhalo \ \citep[e.g.,][]{Klypin2011}. Since the halo rotation curve (RC) is fairly flat near its maximum, galaxies with flat outer RCs provide a good estimate of the host halo's peak velocity, \vflat$\approx$\Vhalo, and consequently of the halo's mass. As a result, the \vflat-based BTFR is believed to be the best representation of the ``fundamental'' relation, and is found by many studies to have lower scatter than the linewidth-based BTFR \citep[e.g.,][]{Verheijen2001,McGaugh2012}.

\begin{figure}[htb]
\centering
\includegraphics[scale=0.47]{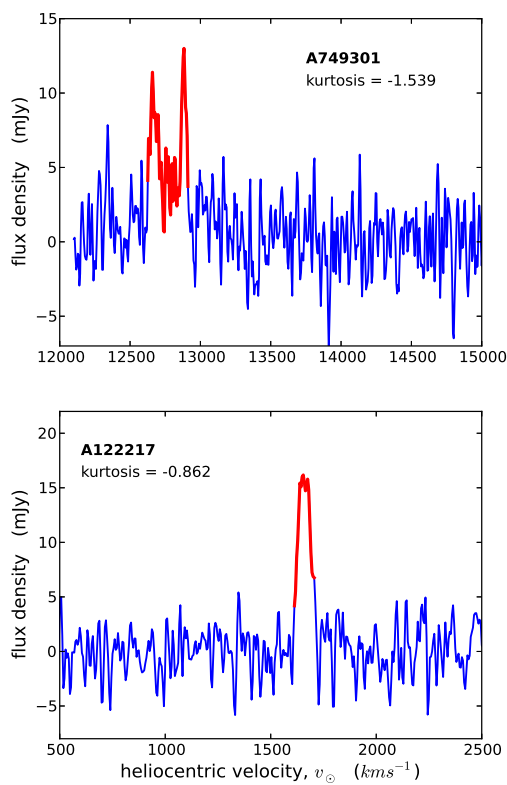}
\caption{ 
Shape measurement for HI line profiles. The figure shows the ALFALFA spectrum for the galaxy with the lowest (top panel) and highest (bottom panel) kurtosis in the \gdtf\ sample. The parts of the spectrum that have been used for measuring the kurtosis are highlighted with a thick red line. Profile kurtosis is a measure of the ``peakiness'' of the profile, and so the top and bottom panels respectively represent the most ``double-horned'' and the most ``peaked'' profiles among the \gdtf \ galaxies.      
 }
\label{fig:kurt_examples}
\end{figure}

Measuring the value of \vflat \ requires spatially resolved HI kinematics, which can only be obtained with interferometric HI observations. By contrast, the sample of galaxies used in this article has been selected from the dataset of the single-dish ALFALFA survey, and therefore only possesses spatially unresolved 21cm data. Of course, using single-dish HI data for measuring the BTFR comes with a very important observational advantage, namely the fact that such data are available for many thousands of objects detected blindly by large HI surveys (such as ALFALFA). Indeed, using the ALFALFA catalog has enabled us to define the \gdtf \ sample in the first place, since heavily gas-dominated galaxies are an intrinsically very rare population. At the same time however, the \vrot \ values obtained from the width of spatially unresolved HI profiles do not necessarily correspond to \vflat, a fact that complicates the theoretical interpretation of the BTFR measurement (see Sec. \ref{sec:discussion}). 

In the context of the \gdtf \ sample, the most important concern is related to the fact that galaxies at the low-end of the mass range probed by the sample may have HI disks that are not extended enough to reach the flat part of the RC. In these cases, the rotational velocity derived from the profile width tends to underestimate the asymptotic flat velocity, \vrot$<$\vflat. This effect can cause low-mass galaxies to ``move'' towards lower velocities than what expected based on an extrapolation of the BTFR of more massive objects. This effect can be very pronounced for some objects, in which case they appear as low-side outliers to the BTFR (see Fig. \ref{fig:btfr} in this article or Fig. 6 in \citealp{Zaritsky2014}). Overall, the effect described above can result in a systematic bias, whereby the linewidth-based BTFR can have a shallower slope (and perhaps larger scatter) than the BTFR constructed with \vflat \ values \citep{McGaugh2012}.

In this section, we use the shape of the spatially unresolved HI profile provided by ALFALFA to broadly assess whether the observational systematic described above could be affecting our measurement of the BTFR. In particular, galaxies with ``double horned'' HI profiles typically correspond to galaxies with flat outer RCs, whereby most of the HI emission is found in the frequency channels corresponding to $\pm$\vflat. On the other hand, galaxies with ``single-peaked'' profiles are more likely to have RCs that are rising out to their last measured point, while intermediate cases display ``boxy'' profiles. Accordingly, we decide to quantify the ``peakiness'' of the HI profile of each \gdtf \ galaxy by measuring the kurtosis\footnotemark{}, $k_4$, of their HI line profile. Figure \ref{fig:kurt_examples} illustrates the process of profile kurtosis measurement for two examples galaxies: The top panel shows the spectrum of AGC749301, which has the most ``double-horned'' (i.e., lowest kurtosis) profile in the sample. The bottom panel shows the spectrum of AGC122217, which has instead the most ``peaked'' (i.e., highest kurtosis) profile. Note that even the most peaked profile among the \gdtf \ galaxies has a negative value of kurtosis; this means that all galaxies in our sample have profile shapes with a broader central peak and weaker wings than a pure Gaussian profile.

\footnotetext{By the term ``kurtosis'' we refer here to the excess normalized kurtosis, $k_4$. More specifically, we calculate the fourth central moment of the galactic spectrum, over the velocity range of interest: $M_4 = \int_{v_{min}}^{v_{max}} F_{HI}(v) \cdot (v-\bar{v})^4 \: dv \; / \int_{v_{min}}^{v_{max}} F_{HI}(v) \: dv$. Here, we denote by $F_\mathrm{HI}(v)$ the HI flux density as a function of heliocentric velocity, and by $\bar{v}$ the average velocity calculated as $\bar{v} = \int_{v_{min}}^{v_{max}} v \cdot F_{HI}(v) \: dv \; / \int_{v_{min}}^{v_{max}} F_{HI}(v) \: dv$. We then calculate the normalized kurtosis, $m_4$, by dividing the fourth central moment by the square of the spectrum's variance: $m_4 = M_4 / \sigma^4$. The variance is the second central moment of the profile, $\sigma^2 = \int_{v_{min}}^{v_{max}} F_{HI}(v) \cdot (v-\bar{v})^2 \: dv \; / \int_{v_{min}}^{v_{max}} F_{HI}(v) \: dv$. Lastly, we compute the excess normalized kurtosis as $k_4 = m_4 - 3$. According to this definition, a Gaussian profile has $k_4 = 0$. Profiles with $k_4 < 0$ are ``platykurtic'', which means they have a broader central peak and weaker wings than a Gaussian profile.}

\begin{figure}[htb]
\centering
\includegraphics[width=\linewidth]{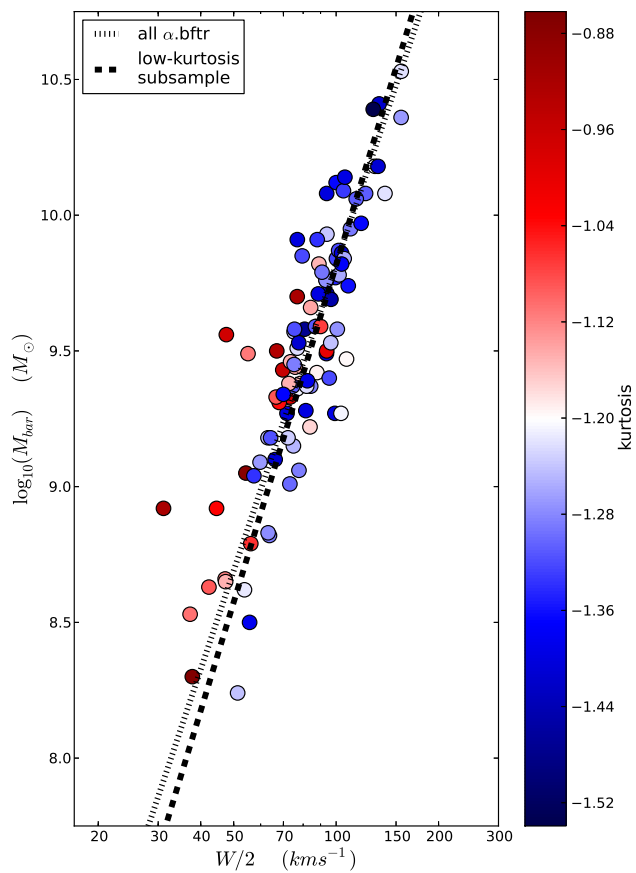}
\caption{ 
BTFR color-coded by HI profile shape. 
The datapoints are the same as in Fig. \ref{fig:btfr} but are color-coded according to the kurtosis of their HI line profile (see \S\ref{sec:profile_shape}). Errorbars have been omitted for clarity. Shades of red correspond to high-kurtosis (``peaked'') profiles, while shades of blue correspond to low-kurtosis (``double-horned'') profiles. The dotted line is the linear fit to the overall \gdtf \ sample (same as in Fig. \ref{fig:btfr}), while the dashed line is a fit to the low-kurtosis subsample of \gdtf \ galaxies with $k_4 < -1.20$. The corresponding slopes are $\alpha = 3.75 \pm 0.11$ and $4.13 \pm 0.15$, respectively.
}
\label{fig:btfr_kurt}
\end{figure}

Figure \ref{fig:btfr_kurt} shows the placement of \gdtf \ galaxies on the BTFR diagram, depending on the kurtosis of their HI profile. Shades of blue represent galaxies with more double-horned profiles, while shades of red correspond to galaxies with more peaked profiles. In general, red shaded datapoints become more frequent as one moves to lower masses along the BTFR. This trend is expected, since the RCs of less luminous galaxies have steeper outer slopes on average \citep[e.g.,][]{Spekkens2005,Catinella2006,Swaters2009}.  However, the most interesting trend we observe in Fig. \ref{fig:btfr_kurt} is the tendency of \gdtf \ galaxies with peaked profiles to fall on the low-velocity side of the best fit to the BTFR. 
Figure \ref{fig:btfr_kurt} seems therefore to support the results of literature studies that find a steeper slope for the \vflat-based relation compared to the linewidth-based relation. More specifically, if we restrict ourselves to the sample of \gdtf \ galaxies with $k_4 < -1.20$ (i.e., those galaxies that are more likely to have flat outer RCs), then the fitted BTFR slope becomes steeper, $\alpha^{(\mathrm{low-}k)} = 4.13 \pm 0.15$. Keep in mind that despite the large numerical difference between the slope values measured for the full \gdtf \ sample and for the low-kurtosis subsample, the fits themselves are not very different (see Fig. \ref{fig:btfr_kurt}). In fact, the formal statistical difference between the two slopes is at the $\approx 2\sigma$ level, and therefore the significance is marginal.



\section{Comparisons with previous work and with theoretical expectations}   
\label{sec:discussion}

\subsection{Comparison with previous work}
\label{sec:btfr_vs_others}

\begin{figure}[htb]
\centering
\includegraphics[scale=0.41]{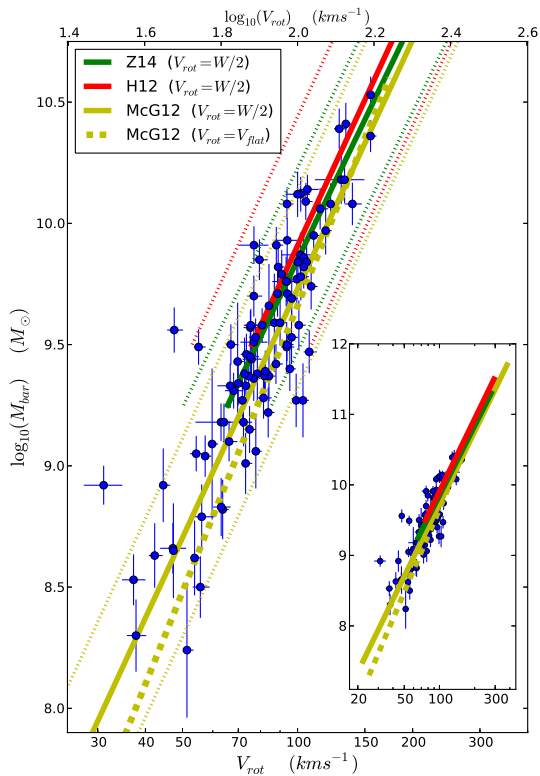}
\caption{ 
Comparison of the BTFR measured here to literature results.
\textit{main figure}: Blue datapoints with errorbars are the galaxies in the \gdtf \ sample (same as in Fig. \ref{fig:btfr}). The green, red, and yellow solid lines are linear fits to the BTFR as reported by \citet[Z14]{Zaritsky2014}, \citet[H12]{Hall2012}, and \citet[McG12]{McGaugh2012} respectively. The dotted colored lines denote the $\pm 2\sigma$ observed scatter of galaxies about the corresponding fit. All three literature results are directly comparable to the \gdtf \ measurement, since they all refer to linewidth-derived rotational velocities (\vrot$=W/2$). The yellow dashed line corresponds to the BTFR measured in terms of \vflat \ by \citet{McGaugh2012}, based on a subsample of their galaxies with spatially resolved HI kinematics. 
All literature fits to the BTFR extend over the range in baryonic mass probed by the corresponding measurement.
\textit{inset panel}: Same as in the main figure but zoomed-out, in order to illustrate the full range of baryonic mass covered by the three literature studies.    
}
\label{fig:btfr_vs_others}
\end{figure}

Figure \ref{fig:btfr_vs_others} compares the BTFR presented in Fig. \ref{fig:btfr} with previous measurements in the literature. First, we consider the measurements of \citet{Zaritsky2014} and \citet{Hall2012}, referred to as Z14 and H12 respectively. Both measurements are based on optically selected samples, and cover the range of intermediate to high baryonic masses (\mbar \ $ \approx 10^{9.5} - 10^{11.5} \; M_\odot$). In particular, the Z14 measurement is based on the galaxy sample of the Spitzer Survey of Stellar Structure in Galaxies (S$^4$G;  \citealp{Sheth2010}). Stellar masses are estimated from 3.6\micron \ and 4.5\micron \ \textit{Spitzer} photometry, according to the calibration of \citet{Eskew2012}. Moreover, Z14 include an estimate of the molecular gas mass in their calculation of baryonic mass. Since the molecular gas fraction is higher for earlier-type galaxies \citep[e.g.,][]{McGaughdeBlok1997}, the inclusion of $M_\mathrm{H_2}$ in the BTFR can lead to a very slight steepening of the slope. The H12 measurement, on the other hand, is based on a sample of spiral galaxies with available SDSS photometry. Stellar masses in H12 are calculated from SDSS $g$ and $i$ band photometry, according to the calibration of \citet{BelldeJong2001}. Similarly to our measurement, H12 do not include molecular gas in their estimate of baryonic mass.

Both the Z14 and H12 measurements use linewidth-derived rotational velocities (\vrot$=W/2$), and therefore can be straightforwardly compared to our measurement of the BTFR.
Unlike in the case of the \gdtf \ sample however, the Z14 and H12 samples consist predominantly of star-dominated galaxies (especially at the high mass end).
Figure \ref{fig:btfr_vs_others} shows that both measurements are in fairly good agreement with the BTFR measured in this article, over the mass range of overlap. 
The BTFR slope measured by Z14 and H12 is in the range $\alpha = 3.3 - 3.4$, which is somewhat shallower than the value of $\alpha = 3.75 \pm 0.11$ measured in this article.
As far as the scatter of galaxies about the best fit line is concerned, the value reported by Z14 agrees very well with that obtained in this work, while the value reported by H12 is somewhat higher.

Second, we consider the BTFR measurement of \citet{McGaugh2012}, hereafter referred to as McG12. The McG12 measurement extends over a very wide range of baryonic mass (\mbar$\approx 10^7 - 10^{11.5} \; M_\odot$), thanks to the inclusion of both star-dominated galaxies covering the high-mass end of the BTFR, and gas-rich galaxies covering the low-mass end of the relation. The gas-rich subsample of the McG12 galaxies\footnotemark{} have also measurements of their spatially resolved RCs, such that McG12 can measure the BTFR both with linewidth-derived velocities and with \vflat \ values. Figure \ref{fig:btfr_vs_others} shows that the linewidth-based\footnotemark{} BTFR measured by McG12 is also in fairly good agreement with the BTFR measured in this article. McG12 finds a slope of $\alpha = 3.41 \pm 0.08$, which is again somewhat shallower than what measured here. 
The total scatter of the linewidth-based BTFR is found by McG12 to be $\sigma_\perp \approx 0.06$~dex, a value very similar to the one found in this work. 

On the other hand, the \vflat-based BTFR of McG12 displays a steeper slope of $\alpha = 3.94 \pm 0.11$. This latter slope value is consistent with the one obtained for the subsample of \gdtf \ galaxies with low profile kurtosis.
Since these galaxies are the most likely objects to have $W/2 \approx$ \vflat \ (refer to \S\ref{sec:profile_shape}), we conclude that the \vflat-based BTFR measured in McG12 is broadly consistent with the \gdtf \ sample, as well.

\footnotetext{
Compared to the McG12 sample, the \gdtf \ sample has two important observational advantages: First, the \gdtf \ galaxies are significantly more gas-rich than the McG12 galaxies ($M_\mathrm{gas}/M_\ast \gtrsim 2.7$ vs. $M_\mathrm{gas}/M_\ast \gtrsim 1$, respectively). Consequently, the uncertainty in stellar mass estimates is much less of an issue for \gdtf \ galaxies than for the McG12 gas-rich galaxies. Second, the \gdtf \ sample is comprised by 97 galaxies, which is about a factor of two more than the 47 gas-rich galaxies in the McG12 sample. At the same time, galaxies in the McG12 sample have the advantage of possessing measurements of their spatially resolved HI kinematics, and of having more accurate distance estimates than the \gdtf \ galaxies (based on TRGB , Cepheids, etc.).
}

\footnotetext{Note that profile widths in McG12 are defined at the 20\% of the peak flux level ($W_{20}$).}

\subsection{Comparison with semi-analytic models in \LCDM}
\label{sec:btfr_vs_sams}

The \LCDM \ cosmological model makes no a priori prediction regarding the  properties of the BTFR. Under a set of very basic assumptions --namely, that the baryon fraction for all halos is constant and that the rotational velocity measured observationally is approximately equal to the peak halo velocity-- the BTFR is expected to follow a power-law of the form $M_\mathrm{bar} \propto V_\mathrm{rot}^3$. This form simply reflects the tight correlation between the mass of a halo and its maximum rotational velocity observed in DM simulations, $M_h \propto V_\mathrm{h,max}^3$ \citep[e.g.,][]{Klypin2011}.

\begin{figure*}[htb]
\centering
\includegraphics[scale=0.39]{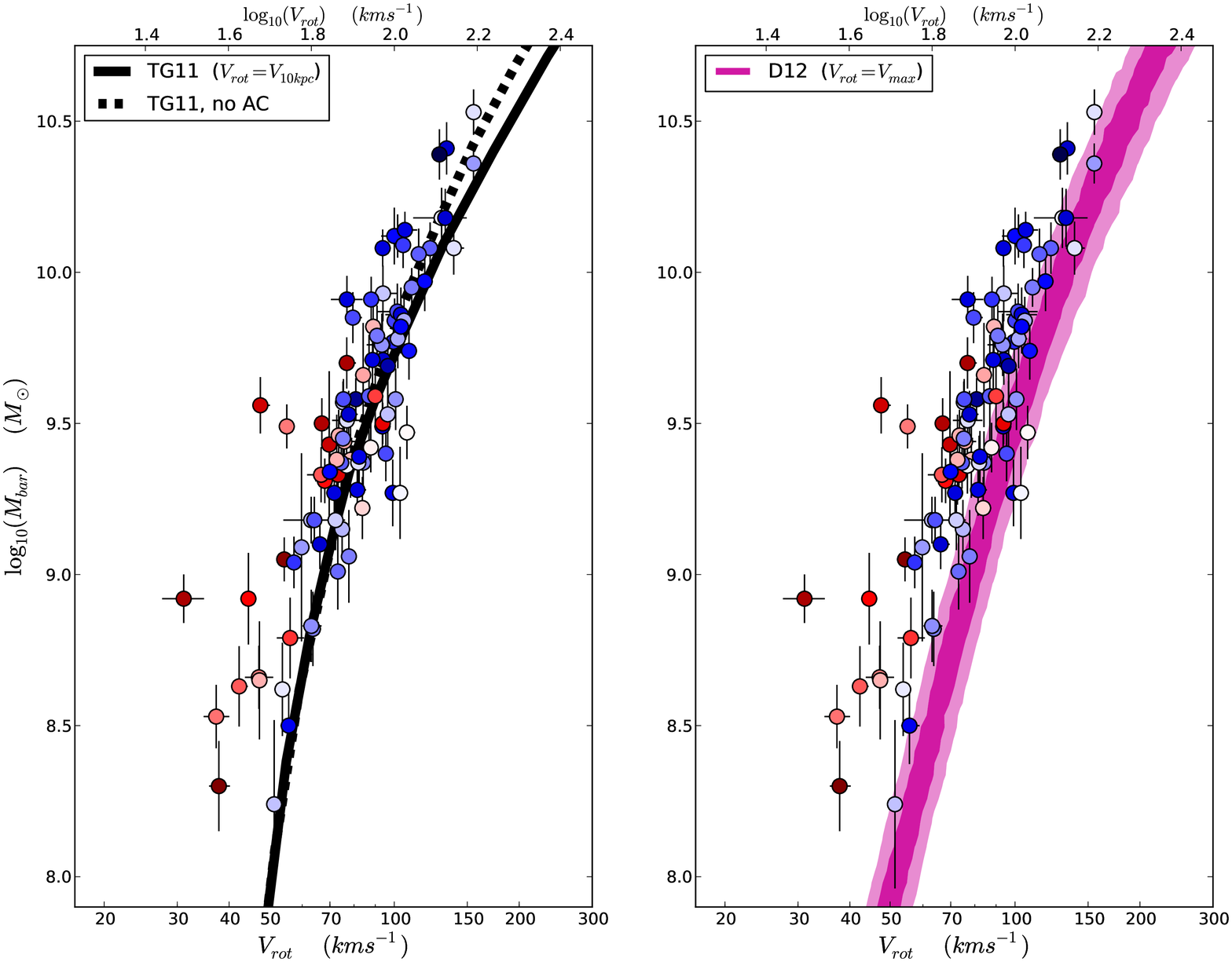}
\caption{ 
Comparison of the BTFR to semi-analytic models in \LCDM.
\textit{left panel}: Datapoints with errorbars are the galaxies in the \gdtf \ sample, color-coded by the kurtosis of their HI profile (same as in Fig. \ref{fig:btfr_kurt}). The black solid line is the prediction for the BTFR based on the semi-analytic model of \citet{Trujillo2011}. The black dashed line represents the variant of the TG11 model that does not include adiabatic contraction of the dark halo due to the infall of baryons. \textit{right panel}:  The magenta shaded region corresponds to the prediction for the shape and intrinsic scatter of the BTFR, according to the semi-analytic model of \citet[D12]{Desmond2012}. Keep in mind that in both panels the rotational velocity of the model is defined in a different way than that for the \gdtf \ galaxies (refer to \S\ref{sec:btfr_vs_sams}). 
}
\label{fig:btfr_vs_sams}
\end{figure*}

A more refined calculation of the BTFR expected in \LCDM \ can be made using semi-analytic models of galaxy formation. These models are typically calibrated to reproduce certain features of the galactic population, for example the measured stellar mass function of galaxies. Semi-analytic models can therefore be used to assign realistic (i.e., observationally motivated) values of stellar and gas mass to halos. The values of $M_\ast$ and $M_\mathrm{gas}$ assigned to a halo by a given model directly determine the $y$-axis position of the halo on the BTFR diagram. Placing a halo on the $x$-axis of the diagram is slightly more complicated: First, a realistic RCs need to be computed, by adding the velocity contribution of the baryonic components to the RC of each DM halo (see for example Fig. 5 in \citealp{Trujillo2011}). This step requires not only model estimates of $M_\ast$ and $M_\mathrm{gas}$, but also of the radial profile of each baryonic component. Lastly, the model needs to report its results in terms of some measure of rotational velocity, which is in effect a ``one-value summary'' of the full simulated RC.

Figure \ref{fig:btfr_vs_sams} compares the BTFR measured for the \gdtf \ sample with the predictions of two semi-analytic \LCDM \ models. The first is the model of \citet{Trujillo2011}, referred to here as TG11. In this model, halos are assigned stellar masses based on an abundance matching (AM) relation, which reproduces the stellar mass function of galaxies measured by \citet{LiWhite2009}. Halos are then assigned gas masses based on an empirical scaling relation between stellar mass and gas fraction \citep{Baldry2008}. TG11 further consider two variants of their model: one where DM halos undergo adiabatic contraction (AC) due to the infall of baryons towards the halo center, and one without AC. Lastly, they report their result in terms of the rotational velocity of their simulated RCs at a fixed radius of 10 kpc for all halos ($V_{\rm rot} = V_{\rm 10kpc}$).

The left panel of the figure shows that the TG11 model produces a curved BTFR, regardless of whether or not AC is included in the modeling. Curvature in the BTFR is in fact a generic prediction of \LCDM \ semi-analytic models, which stems from the need to achieve low baryon fractions for both high-mass and low-mass halos \citep[see e.g.,][Fig. 17]{Papastergis2012}. Overall, the BTFR predicted by the TG11 model without AC displays mild enough curvature, such that it cannot be ruled out by the \gdtf \ data. This is especially true when we consider the \gdtf \ galaxies with low values of profile kurtosis (blue shaded points); these datapoints can be compared to the model more directly, since the corresponding galaxies are more likely to have $W/2 \approx V_{\rm 10kpc}$.

The right panel of Fig. \ref{fig:btfr_vs_sams} compares instead the \gdtf \ measurement of the BTFR with the semi-analytic model of \citet{Desmond2012}, referred to here as D12. This model also uses an AM relation to assign stellar masses to halos, in particular the one for late-type galaxies derived by \citet{Rodriguez2011}. Gas masses are assigned once again based on empirical scaling relations between stellar mass and gas fraction \citep{Avila2008,McGaugh2005}. Lastly, the D12 model prediction for the BTFR is expressed in terms of the peak velocity of the simulated RCs ($V_{\rm rot} = V_{\rm max}$). One unique feature of the D12 model is that, in addition to modeling the average BTFR, it also explicitly models the intrinsic scatter of the relation expected in \LCDM. In particular, it accounts for three sources of intrinsic scatter:
(i) Scatter in the concentration parameter ($c$) of halos, (ii) scatter in the spin parameter ($\lambda$) of halos, and (iii) scatter in the galactic baryon fraction ($f_\mathrm{bar} = M_\mathrm{bar}/M_h$).

Figure \ref{fig:btfr_vs_sams} shows that the main discrepancy between the D12 model and the \gdtf \ measurement is a systematic shift of the model with respect to the data. It is unclear whether this shift constitutes a genuine shortcoming of the D12 model, since it could be due to either a systematic overestimate of \vrot \ at fixed \mbar, or to a systematic underestimate of \mbar \ at fixed \vrot. 
In this latter case, there is a possibility that the shift reflects a systematic offset in the AM result used by D12 to assign baryonic masses to halos.

On the other hand, the shape of the BTFR is a more robust prediction of a semi-analytic model.
The D12 model predicts a relation with little curvature, which seems to reproduce well the observed shape of the relation of the \gdtf \ sample.     
Lastly, the intrinsic scatter of the relation in \LCDM \ according to the D12 model is $\sigma_{\perp,\mathrm{intr}} \approx 0.045$~dex (in the direction perpendicular to the mean relation). This value is compatible with the upper limit on the intrinsic scatter of the BTFR derived from the \textit{total} scatter of the \gdtf \ galaxies (refer to Sec. \ref{sec:btfr}).

\subsection{Comparison with hydrodynamic simulations in \LCDM}
\label{sec:btfr_vs_sims}

\begin{figure}[htb]
\centering
\includegraphics[scale=0.41]{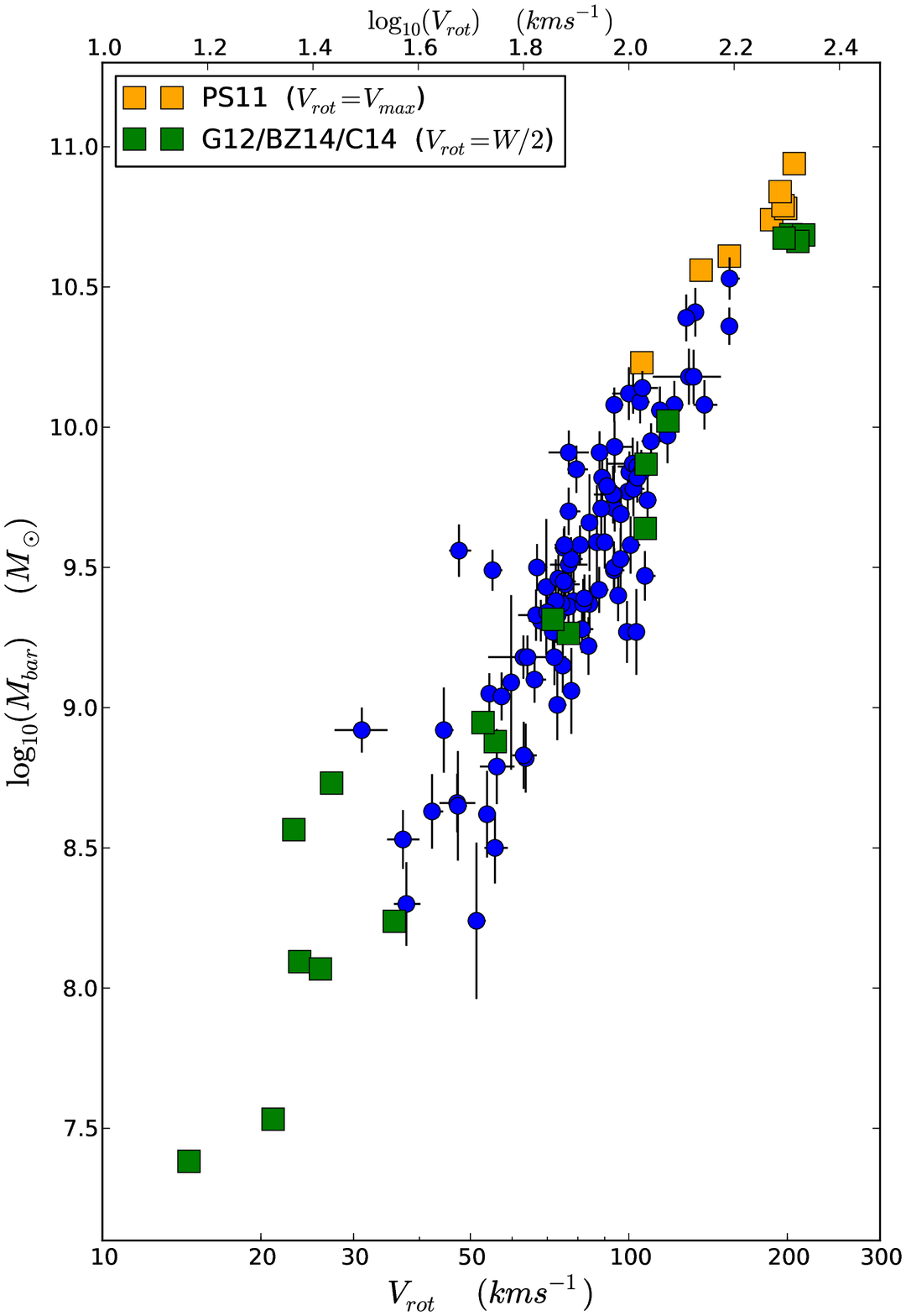}
\caption{ 
Comparison of the BTFR to hydrodynamic simulations in \LCDM.
Blue datapoints with errorbars are the galaxies in the \gdtf \ sample (same as in Fig. \ref{fig:btfr}). The orange squares are mock galaxies from the hydrodynamic simulation of \citet{PiontekSteinmetz2011}, while the green squares are mock galaxies produced in a set of hydrodynamic simulations by \citet[G12]{Governato2012}, \citet[BZ14]{BrooksZolotov2014}, and \citet[C14]{Christensen2014}. Note that rotational velocities for the G12/BZ14/C14 mock galaxies are reported in the same way as for the \gdtf \ galaxies, i.e., in terms of the linewidth of their simulated HI profiles.
}
\label{fig:btfr_vs_sims}
\end{figure}

Figure \ref{fig:btfr_vs_sims} compares the BTFR measured for the \gdtf \ sample with a set of mock galaxies produced in hydrodynamic simulations in the \LCDM \ context. In particular, we consider eight mock galaxies created in the simulation of \citet{PiontekSteinmetz2011}, denoted as PS11. We also consider eighteen mock galaxies produced in a set of hydrodynamic simulations  by \citet{Governato2012}, \citet{BrooksZolotov2014}, and \citet{Christensen2014}, denoted as G12/BZ14/C14. The PS11 mock galaxies cover the high end of the mass range probed by the \gdtf \ sample (\mbar$ \approx 10^{10} - 10^{11} \; M_\odot$), while the G12/BZ14/C14 simulations contain mock galaxies with a very wide range in baryonic mass (\mbar$ \approx 10^{7.5} - 10^{10.5} \; M_\odot$). One important point to keep in mind regarding the G12/BZ14/C14 mock galaxies is that their rotational velocities are they are extracted from simulated HI profiles (similar to those shown in the third column of Fig. \ref{fig:selection}). This means that the location of these mock galaxies on the BTFR can be directly compared to the location of the \gdtf \ galaxies. Note also that we plot only ``central'' mock galaxies from this latter simulation, i.e., we exclude mock galaxies hosted by subhalos.

Figure \ref{fig:btfr_vs_sims} shows that the hydrodynamic simulations considered in this article are successful at reproducing the properties of the BTFR, over most of the range of masses probed by the \gdtf \ sample (\mbar$ \approx 10^{9} - 10^{10.5} \; M_\odot$). At slightly higher masses the agreement is also presumably good, since both simulation works are compatible with an extrapolation of the relation measured for the \gdtf \ sample. 
The situation is not as straightforward at the low-mass end of the BTFR, however. In particular, mock galaxies in the G12/BZ14/C14 simulations with \mbar$< 10^9 \; M_\odot$ show an abrupt increase in the scatter of their BTFR. In addition, low-mass mock galaxies tend to be located at lower rotational velocities than what an extrapolation of the \gdtf \ measurement to low masses would suggest. Unfortunately, the \gdtf \ sample contains only a handful of objects with \mbar$< 10^9 \: M_\odot$, and so it cannot be used to thoroughly test the behavior predicted by this simulation set at low masses. 
As a result, this predicted abrupt change in the scatter and normalization of the linewidth-based BTFR at low masses (\mbar$< 10^9 \; M_\odot$) should be tested against samples specifically targeting low-mass dwarfs \citep[see, e.g., recent work by][]{Sales2016}.

\subsection{Comparison with the predictions of MOND}
\label{sec:btfr_vs_mond}

\begin{figure}[htb]
\centering
\includegraphics[scale=0.41]{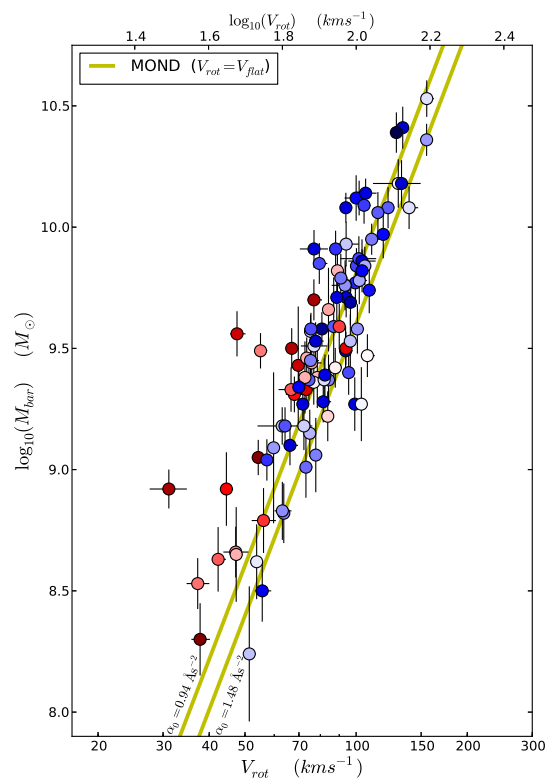}
\caption{ 
Comparison of the BTFR to the MOND prediction.
Datapoints with errorbars are the galaxies in the \gdtf \ sample, color-coded by the kurtosis of their HI profile (same as in Fig. \ref{fig:btfr_kurt}). The two yellow solid lines represent the MOND prediction for the BTFR \citep{Milgrom1983b}, for two values of the theory's acceleration parameter, $\alpha_0$ (indicative of the parameter's measurement uncertainty; see \S\ref{sec:btfr_vs_mond} for details). 
MOND further predicts that the BTFR is a perfect power-law, i.e., it predicts a relation with zero intrinsic scatter. Keep in mind that the MOND prediction refers to asymptotic flat velocities (\vrot$=$\vflat), while rotational velocities for \gdtf \ galaxies are linewidth-derived values (\vrot$= W/2$).
}
\label{fig:btfr_vs_mond}
\end{figure}

Unlike in the case of \LCDM, MOND makes a strong a priori prediction regarding the shape and intrinsic scatter of the BTFR \citep{Milgrom1983b}. In particular, MOND predicts that the relation has a slope of exactly 4 when the rotational velocity is expressed in terms of the asymptotic flat velocity, $M_{\rm bar} \propto V_{\rm flat}^4$. Furthermore, the normalization of the BTFR depends only on the value of the theory's acceleration parameter, $\alpha_0$ (refer to \S3.2 in \citealp{McGaugh2012} for a derivation of the properties above). Lastly, in a MOND universe there is no dark matter, and therefore gravitational forces are entirely determined by the amount of baryonic matter present in a galaxy. As a result, MOND predicts that a single power law relation with \textit{zero} intrinsic scatter is sufficient to describe the \vflat-based BTFR of all types of galaxies. 

Figure \ref{fig:btfr_vs_mond} tests the MOND prediction for the BTFR against the observed relation for the \gdtf \ galaxies. Keep in mind that while the slope and intrinsic scatter of the MOND prediction are fixed, the predicted normalization can vary slightly, due to the uncertainty in the observational determination of $\alpha_0$ (e.g., $\alpha_0 = 1.21 \pm 0.27 \: \mathrm{\AA} \mathrm{s}^{-2}$ according to \citealp{Begeman1991}). The main complication in interpreting the comparison of Fig. \ref{fig:btfr_vs_mond} is the fact that rotational velocities for the \gdtf \ galaxies are linewidth-derived values ($W/2$), but the MOND predictions refer to flat values (\vflat). However, focusing our attention to galaxies with low profile kurtosis can help us perform a comparison of the theory's predictions against the objects that are more likely to have $W/2 \approx V_\mathrm{flat}$ (refer to \S\ref{sec:profile_shape}). Figure \ref{fig:btfr_vs_mond} shows that, as far as the shape of the BTFR is concerned, the MOND prediction is compatible with the relation measured for the low-kurtosis subsample of \gdtf \ galaxies; this is also reflected in the fact that the fitted slope to the low-kurtosis subsample is $\alpha^{\mathrm{(low-}k\mathrm{)}} = 4.13 \pm 0.15$, a value that is compatible with the MOND prediction of $\alpha = 4$. The normalization of the observed BTFR implies an acceleration parameter value of $\alpha_0 = 0.94 \; \mathrm{\AA s}^{-2}$, again compatible with the measurement of \citet{Begeman1991}. We would like here to remind the reader that the best fit slope can be different among different fitting methods, and this can affect the assessment of the consistency between our sample and the MOND prediction (see Appendix \ref{sec:alternative_fits}).

\begin{figure}[htb]
\centering
\includegraphics[scale=0.47]{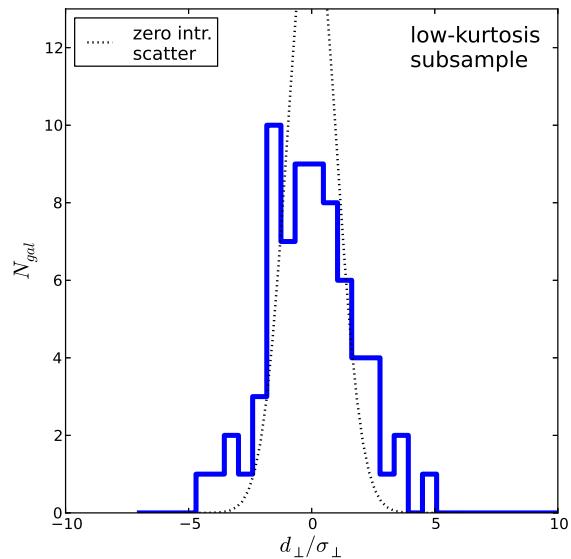}
\caption{ 
Same as Fig. \ref{fig:histogram}, but restricted to the subsample of \gdtf \ galaxies with low profile kurtosis ($k_4 < -1.20$). See \S\ref{sec:btfr_vs_mond} for the scientific interpretation of this figure. 
}
\label{fig:histogram_dh}
\end{figure}

As far as the MOND prediction of zero intrinsic scatter is concerned, we repeat here the test of Fig. \ref{fig:btfr_kurt}, but this time restricting ourselves to the subsample of \gdtf \ galaxies with low profile kurtosis. The result is shown in Figure \ref{fig:histogram_dh}. The figure shows that, similarly to the case of the overall \gdtf \ sample, the low-kurtosis subsample displays a small (but non-zero) value of intrinsic scatter. Once again, the result of Fig. \ref{fig:histogram_dh} is supported by the analysis performed in Appendix \ref{sec:alternative_fits} (see Fig. \ref{fig:chi2_ellipse_scatter_dh}). Taken at face value, the result of Fig. \ref{fig:histogram_dh} is incompatible with the prediction of MOND. However, keep in mind that the level of intrinsic scatter in the linewidth-based BTFR indicated by Fig. \ref{fig:histogram_dh} is rather small, and it could be accounted for by slight mismatches between values of $W/2$ and \vflat \ for our galaxies. Eventually, a more direct comparison between the prediction of MOND and the measured BTFR should be made, with the aid of spatially resolved kinematic data for the \gdtf \ galaxies. Such data will become available in the near future, thanks to a follow-up campaign of HI interferometric observations that has been carried out for several objects in the \gdtf \ sample \citep{Papastergis2016+}.

\section{Summary \& discussion}
\label{sec:summary}

We select galaxies from the ALFALFA 21cm survey \citep{Haynes2011} with properties that are ideal for an accurate measurement of the BTFR. In particular, the selected galaxies have reliable ALFALFA and SDSS data, and most importantly they are gas-dominated ($M_\mathrm{gas}/M_\ast \gtrsim 2.7$) and oriented edge-on. The final sample contains 97 galaxies, which we refer to as the ``\gdtf'' sample (Sec. \ref{sec:dataset}).
Thanks to their high gas fractions, the baryonic content of \gdtf \ galaxies consists mostly of atomic gas. This means that baryonic masses can be measured with high accuracy for this sample, despite the $\sim$60\%-100\% uncertainty inherent in stellar mass estimates (Sec. \ref{sec:derived_quantities}).
Moreover, the edge-on orientation of \gdtf \ galaxies means that their rotational velocity can be calculated straightforwardly from the observed width of their HI lineprofile (\vrot$= W/2$), and therefore the uncertainty associated with inclination corrections is almost eliminated (\S\ref{sec:velocity_width}).

The BTFR measured with our sample of gas-dominated and edge-on ALFALFA galaxies consists of a well-defined ``main body'', with only three clear outliers (see Fig. \ref{fig:btfr}). 
Excluding the outliers, we measure a slope of $\alpha = 3.75 \pm 0.11$, a value that is somewhat steeper than (but in broad agreement with) the typical values obtained in the literature for the linewidth-based BTFR \citep[e.g.,][]{Hall2012,McGaugh2012,Zaritsky2014}. 
The relation is also remarkably tight, with most \gdtf \ galaxies contained within a band that is just 0.2 dex wide in the direction perpendicular to the best fit line (see Fig. \ref{fig:perp+parallel}). 
Despite the tightness of the relation, the small error budget of the \gdtf \ sample enables us to test for the presence of low levels of intrinsic scatter. We find that, according to the \gdtf \ sample, the linewidth-based BTFR has a small (but non-zero) value of intrinsic scatter (see Fig. \ref{fig:histogram}).

We furthermore study how the position of our galaxies on the BTFR diagram depends on the shape of their HI line profile (Fig. \ref{fig:kurt_examples}). We find a systematic trend, whereby galaxies with high kurtosis (i.e., more ``peaked'') profiles are more likely to be located on the low-velocity side of the best fit line than galaxies with low kurtosis (i.e., more ``double-horned'') profiles (see Fig. \ref{fig:btfr_kurt}). Since galaxies with ``double-horned'' profiles are more likely to have a flat outer part in their resolved RCs, we anticipate that the BTFR of the low-kurtosis subsample of \gdtf \ galaxies should be more representative of the BTFR expressed in terms of \vflat. When we restrict the \gdtf \ sample to objects with low kurtosis profiles, we measure a slightly steeper slope for the relation, $\alpha = 4.13 \pm 0.15$.

Overall, the linewidth-based BTFR presented in this article is intended to be a reliable observational benchmark, against which to test theoretical expectations. 
In this article, we compare our measurement with a representative set of semi-analytic models and hydrodynamic simulations in the \LCDM \ context \citep{Trujillo2011,Desmond2012,PiontekSteinmetz2011,Governato2012,BrooksZolotov2014,Christensen2014}, as well as with MOND \citep{Milgrom1983b}.

According to semi-analytic models, the BTFR is expected to be ``curved'' in a \LCDM \ universe (refer to \S\ref{sec:btfr_vs_sams}). However, the predicted curvature is small enough that it cannot be ruled out by the \gdtf \ sample alone (see Fig. \ref{fig:btfr_vs_sams}). In this respect, measuring the BTFR over a very extended range in baryonic mass can offer more stringent constraints on \LCDM \ semi-analytic models. In such a test, the \gdtf \ sample can act as a reliable anchor point bridging measurements at the high-mass end of the BTFR \citep[e.g.,][]{NoordermeerVerheijen2007,denHeijer2015} with measurements at the low-mass end. As far as the intrinsic scatter of the BTFR is concerned, the prediction from semi-analytic modeling is probably consistent with the observed intrinsic scatter of the \gdtf \ sample (see right panel of Fig. \ref{fig:btfr_vs_sams}).

Hydrodynamic simulations in \LCDM \ seem to reproduce the observed BTFR slightly better compared to semi-analytic models. For example, the mock galaxies produced in the \citet{PiontekSteinmetz2011} and \citet{Governato2012,BrooksZolotov2014,Christensen2014} simulations show very little curvature in their BTFR over 3.5 dex in baryonic mass. In addition, the BTFR produced in the latter set of simulations shows remarkably small intrinsic scatter at intermediate masses, matching well the level of intrinsic scatter indicated by the \gdtf \ sample. The fact that hydrodynamic simulations can achieve lower levels of intrinsic scatter compared to semi-analytic models may point to the fact that certain galaxy properties that are related to the intrinsic scatter of the BTFR (e.g., halo spin, halo concentration, galactic baryon fraction) may not be entirely independent of each other. A further feature of the G12/BZ14/C14 set of simulations is that, at low baryonic masses (\mbar$ < 10^{9} \; M_\odot$), the scatter in the BTFR increases considerably and galaxies are positioned on the low-velocity side of the BTFR as extrapolated from the \gdtf \ sample measurement (see Fig. \ref{fig:btfr_vs_sims}). This predicted behavior can be tested by combining the \gdtf \ sample covering intermediate masses with samples that probe the BTFR in the dwarf galaxy regime \citep[e.g.,][]{Cote2000,Begum2008a,Trachternach2009,Cannon2011,Hunter2012,Kirby2012,Lelli2014}.

As far as MOND is concerned, the theory's predictions concern the \vflat-based BTFR; as a result we test MOND primarily against the subsample of \gdtf \ galaxies with low profile kurtosis (i.e., galaxies with relatively ``double-horned'' HI profiles, which are therefore more likely to have $W/2 \approx$\vflat). The slope of the BTFR measured from the low-kurtosis subsample is $\alpha = 4.13 \pm 0.15$, which is consistent with the theory's prediction of $\alpha = 4$ (see Fig. \ref{fig:btfr_vs_mond}). Keep in mind though that the derived best fit slope depends on the fitting method employed, something that can affect the comparison between our measurement and the MOND prediction (refer to Appendix \ref{sec:alternative_fits}). The normalization of the BTFR implies instead an acceleration parameter of $\alpha_0 =0.94 \: \mathrm{\AA s}^{-2}$, a value that is also consistent with previous determinations in the literature \citep{Begeman1991}. Lastly, MOND predicts a \vflat-based BTFR with zero intrinsic scatter. This prediction does not seem to be supported by the low-kurtosis subsample of \gdtf \ galaxies, which display a small value of intrinsic scatter (see Fig. \ref{fig:histogram_dh}). It is still possible, however, that small mismatches between the values of $W/2$ and \vflat \ for our sample of galaxies could account for the small level of intrinsic scatter observed.

In general, the biggest complication affecting the comparisons between theoretical models and the \gdtf \ measurement comes from differences in the definition of rotational velocity. Our measurement of the BTFR is expressed in terms of the linewidth-derived rotational velocity. This definition is observationally motivated but it cannot be straightforwardly replicated in theoretical models, since it does not refer to any specific location on a galaxy's resolved RC \citep[e.g.,][]{BrookShankar2016}. Strictly speaking, only the G12/BZ14/C14 set of simulations is directly comparable to our measurement, since this theoretical work derives rotational velocities for their mock galaxies based on the width of their simulated HI profiles. Models that define instead their rotational velocity at a large galactocentric radius (or as the peak or asymptotic velocity in the resolved RC) should be more closely comparable with the low-kurtosis subsample of \gdtf \ galaxies (refer to \S\ref{sec:profile_shape}). In the near future, we plan to present an update of the BTFR measurement, based on a subsample of \gdtf \ galaxies that has been targeted with follow-up interferometric HI observations \citep{Papastergis2016+}. The addition of spatially resolved kinematic information to a number of selected \gdtf \ members will lead to improved comparisons to theoretical models presented in \S\ref{sec:btfr_vs_sams}-\ref{sec:btfr_vs_mond}.

\begin{acknowledgements}
We would like to thank the entire ALFALFA collaboration team for observing, flagging, and extracting the catalog of HI galaxies used in this work. We would also like to thank Andrey Kravtsov for his suggestions, which have greatly improved the current article. Lastly, we would like to thank Tom Jarrett for providing resolved WISE photometry for many of our galaxies, and Alyson Brooks for providing yet-to-be-published data on simulated galaxies.\\ 
EP is supported by a NOVA postdoctoral fellowship at the Kapteyn Astronomical Institute. EAKA is supported by grant TOP1EW.14.105, which is financed by the Netherlands Organisation for Scientific Research (NWO). JMvdH acknowledges support from the European Research Council under the European Union's Seventh Framework Programme (FP/2007-2013)/ ERC Grant Agreement nr. 291531. The ALFALFA team at Cornell is supported by U.S. NSF grant AST-1107390 to Martha Haynes and Riccardo Giovanelli and by grants from the Brinson Foundation.
\end{acknowledgements}
%
%
%
%

\appendix

\section{The impact of internal extinction on the stellar mass estimates of \gdtf \ galaxies}
\label{sec:appendix_extinction}

Edge-on galaxies are typically more heavily affected by internal dust extinction than galaxies of intermediate inclination \citep[e.g.,][]{Giovanelli1994}. One could therefore worry that the stellar mass estimates for \gdtf \ galaxies are systematically underestimated, leading to artificially high gas-fractions. At the same time, the \gdtf \ galaxies have relatively low masses and are gas-dominated. In fact, most of them do not exhibit clear signs of dust obscuration upon visual inspection (e.g., dust lanes). As a result, it is not a priori clear if stellar mass estimates for \gdtf \ galaxies are more severely affected by internal extinction than stellar mass estimates for typical galaxies. 

In order to make a more quantitative assessment of the issue above, we measure Balmer decrements for those galaxies in the ``parent'' ALFALFA sample that also have an entry in the NSA catalog (refer to \S\ref{sec:data_sources}). The parent sample consists of ALFALFA galaxies that satisfy selection criteria \ref{item:quality}-\ref{item:oc} only, and so they cover a wide range in galactic inclination, stellar mass, and gas fraction. The Balmer decrement compares the \balmer \ line ratio measured from the optical spectrum of a galaxy to the intrinsic line ratio expected in star-forming HII regions. Assuming a \citet{Calzetti2000} reddening law, the Balmer decrement is related to optical extinction through the relation

\begin{eqnarray}
A_\mathrm{HII}({H_\alpha}) = 6.56 \times \log_{10} \left( \frac{(H_\alpha/H_\beta)_\mathrm{obs}}{(H_\alpha/H_\beta)_\mathrm{intr}} \right) \;\;\; .
\label{eqn:balmer}
\end{eqnarray}

\noindent
In the formula above, $A_\mathrm{HII}({H_\alpha})$ is the extinction that HII regions experience at the wavelength of \halpha \ (656.3 nm), in units of magnitudes. The intrinsic Balmer ratio of HII regions is adopted here to be $(H_\alpha/H_\beta)_\mathrm{intr} = 2.86$, a standard value used in the literature \citep{OsterbrockFerland2005}.

\begin{figure}[htb]
\centering
\includegraphics[width=\linewidth]{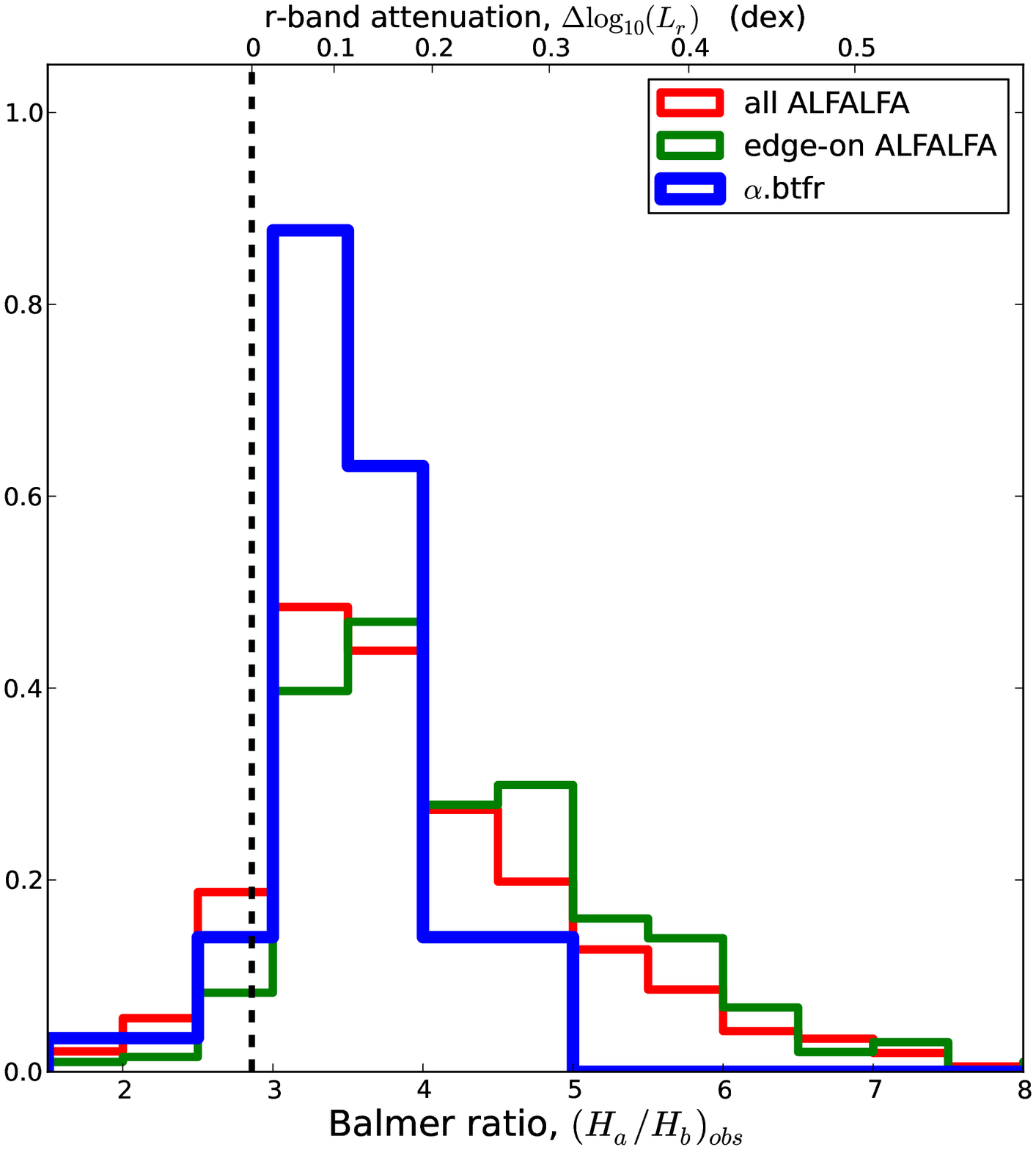}
\caption{ 
Balmer decrements for \gdtf \ galaxies.
The blue thick histogram represents the distribution of Balmer decrements for the galaxies in the \gdtf \ sample. Balmer decrements have been computed from the \halpha \ and \hbeta \ line fluxes reported in the NSA catalog (refer to \S\ref{sec:data_sources}). The red histogram is the Balmer decrement distribution for a broad sample of galaxies that are high quality ALFALFA detections. This ``parent'' sample spans a broad range in galaxy inclination, stellar mass, and gas fraction. The green histogram represents the distribution of the subsample of galaxies in the parent ALFALFA sample that are oriented edge-on. The vertical dashed line denotes an \balmer \ ratio of 2.86, which is adopted here as the fiducial value for HII regions in the absence of extinction. The upper $y$-axis represents the implied dimming of the $r$-band luminosity due to internal extinction (Eqn. \ref{eqn:extinction}).
}
\label{fig:balmer}
\end{figure}

In Figure \ref{fig:balmer} we plot the distribution of Balmer decrements for the overall parent ALFALFA sample, and also for the galaxies in the parent sample that --similarly to the \gdtf \ galaxies-- are oriented edge-on. We see that edge-on galaxies have systematically higher Balmer decrements (indicative of higher internal extinction levels) but not by much. We have verified that this situation arises because Balmer decrements depend primarily on the stellar mass of a galaxy, and only to a lesser extent on its inclination. Figure \ref{fig:balmer} then compares the two distributions described above to the distribution of Balmer decrements for the \gdtf \ galaxies. The figure demonstrates that \gdtf \ galaxies are less affected by internal extinction than the parent  ALFALFA sample. This behavior is the result of the fact that \gdtf \ galaxies tend to have low stellar masses compared to the overall ALFALFA population.

Balmer decrements measure the amount of extinction experienced by HII regions, $A_\mathrm{HII}$. However, the relevant quantity for stellar mass estimates is the extinction experienced by the stellar population, $A_\ast$. Typically, the former extinction measure is higher than the latter, because HII regions are usually embedded in dusty clouds. \citet{Kreckel2013} has shown that the internal extinction levels experienced by these two galactic components are correlated galaxy-wise, following a mean scaling of $A_\ast \approx 0.5 \times A_\mathrm{HII}$ (in units of magnitudes). Since the wavelength of \halpha \ falls within the $r$ optical band, the combination of Eqn. \ref{eqn:balmer} and of the \citet{Kreckel2013} scaling enables us to use observed Balmer decrements of galaxies to infer the attenuation in their $r$-band luminosity caused by internal extinction:

\begin{eqnarray}
\Delta\log_{10}(L_r)  & = & \frac{1}{2.5} \times A_\ast = \frac{1}{2.5} \times 0.5 \times A_\mathrm{HII} \notag \\ 
& = & 1.312 \times \log_{10} \left( \frac{(H_\alpha/H_\beta)_\mathrm{obs}}{2.86} \right) \;\;\; .  
\label{eqn:extinction}
\end{eqnarray}

\noindent
According to Figure \ref{fig:balmer}, the dimming that \gdtf \ galaxies experience due to internal extinction is less than a factor of 1.6~-~2 at $r$-band. Such values are typical among ALFALFA galaxies, and cannot compromise the status of \gdtf \ galaxies as very gas-rich objects in any case.

At the same time, it is important to keep in mind that $\Delta \log_{10}(L_r)$ in Eq. \ref{eqn:extinction} represents a rather conservative upper limit on the true impact of internal extinction on stellar mass estimates. First, Balmer decrements in this work are derived from the spectral line data available in the NSA catalog, which are in turn obtained from post-processing SDSS spectra. These fiber-based spectra cover only the central 3\arcs \ of each galaxy, and tend to sample the dusty central parts of the objects only.

\begin{figure}[htb]
\centering
\includegraphics[scale=0.43]{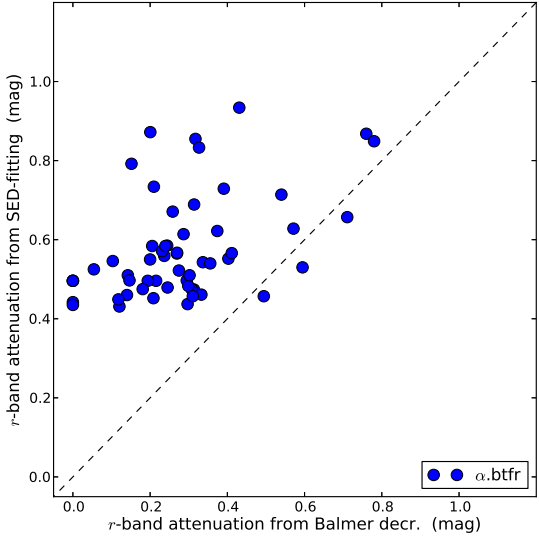}
\caption{ 
Comparison of internal extinction derived from Balmer decrements and SED fitting.
The $x$-axis corresponds to the $r$-band attenuation for the \gdtf \ galaxies implied by their Balmer decrements (Eqn. \ref{eqn:extinction}). The $y$-axis corresponds to the $r$-band attenuation inferred during the process of SED fitting used to calculate $M_\ast^\mathrm{(SDSS)}$ (refer to method \ref{item:sed} in \S\ref{sec:stellar_mass}). The dashed black line is a one-to-one reference line.
}
\label{fig:balmer_vs_sed}
\end{figure}

Second, the dimming effect of internal extinction is fully or partially offset by its reddening effect when computing stellar masses. This is most clearly illustrated in the case of stellar mass estimate $M_\ast^{\mathrm{(}g-i\mathrm{)}}$ (method \ref{item:taylor}): internal extinction tends to decrease the $i$-band apparent magnitude, but at the same time it causes the $g-i$ color to become redder. This means that internal extinction would cause us to apply an overestimated mass-to-light ratio to an underestimated galactic luminosity, therefore producing two mutually offsetting effects.
Stellar mass methods based on SED-fitting (such as methods \ref{item:sed} \& \ref{item:nsa} used in this article) take internal extinction explicitly into account. In particular, these methods include internal extinction as a parameter to be determined during the SED fitting process. Figure \ref{fig:balmer_vs_sed} compares the internal extinction of \gdtf \ galaxies implied by their Balmer decrements (Eqn. \ref{eqn:extinction}) with the internal extinction derived during the estimation of $M_\ast^\mathrm{(SDSS)}$ (method \ref{item:sed}). The figure shows that, if anything, the internal extinction of \gdtf \ galaxies is over-compensated for during the SED fitting process.   
Lastly, stellar mass estimates $M_\ast^\mathrm{(2MASS)}$ and $M_\ast^\mathrm{(WISE)}$  (methods \ref{item:2mass} \& \ref{item:wise}) are based on NIR photometry, and as a result they are minimally affected by dust extinction.


\section{Details of fitting methodology and results of alternative fitting methods}
\label{sec:appendix_fit}

\subsection{Fitting methodology}
\label{sec:fit_details}


The fitting process is performed in logarithmic space. For this reason we define logarithmic variables for the linewidth-derived rotational velocity, $V_\mathrm{rot} = W/2$, and for the baryonic mass, \mbar:

\begin{eqnarray}
v & = &\log_{10}(V_\mathrm{rot} \, / \, \mathrm{km}\, \mathrm{s}^{-1}), \;\;\;\;\; \mathrm{and} \label{eqn:log_vrot} \\
m_b & = & \log_{10}(M_\mathrm{bar}  /  M_\odot) \;\;\; .
\label{eqn:log_mbar}
\end{eqnarray}

\noindent
Accordingly, the errors on the two quantities also need to be expressed in logarithmic units. The error on \mbar \ has already been computed in logarithmic units in \S\ref{sec:baryonic_error} (please follow Eqns. \ref{eqn:mbar_err_1} through \ref{eqn:mhi_err}).

\begin{eqnarray}
\sigma_{m_b} = \sigma_{\log_{10}(M_\mathrm{bar})} \;\;\;\;\; [\mathrm{Eqns.} \; \ref{eqn:mbar_err_1} - \ref{eqn:mhi_err}] \;\;\; .
\label{eqn:log_mbar_err}
\end{eqnarray}

\noindent
The error on \vrot \ is instead reported in terms of linear units in \S\ref{sec:velocity_width} (Eqns. \ref{eqn:width_error_low} - \ref{eqn:vel_error}). We therefore convert it to logarithmic units as follows:

\begin{eqnarray}
\sigma_{v+} & = & \log_{10}(V_\mathrm{rot} + \sigma_{V_\mathrm{rot}+}) - \log_{10}(V_\mathrm{rot}) \;\;\; \mathrm{and} \label{eqn:log_vel_err_low}\\
\sigma_{v-} & = & \log_{10}(V_\mathrm{rot}) - \log_{10}(V_\mathrm{rot} - \sigma_{V_\mathrm{rot}-}) \;\;\; . \label{eqn:log_vel_err_high}
\end{eqnarray}

\noindent
A complication regarding the \vrot \ errors is that they are non-symmetric. In order to simplify the fitting process, we use a symmetrized version of the velocity errors (in logarithmic units), calculated as follows:

\begin{eqnarray}
\sigma_v & = & \frac{1}{2} \: (\sigma_{v+} + \sigma_{v-}) \;\;\; . 
\label{eqn:log_vel_err_symm}
\end{eqnarray}

\noindent
By substituting Eqns. \ref{eqn:log_vel_err_low} \& \ref{eqn:log_vel_err_high} into Eqn. \ref{eqn:log_vel_err_symm}, and by assuming that fractional velocity errors are small ($\sigma_{V_\mathrm{rot}\pm} \ll V_\mathrm{rot}$), we obtain after some algebraic manipulation:

\begin{eqnarray} 
\sigma_v & = & \frac{1}{2} \: \left( \:\log_{10}(V_\mathrm{rot} + \sigma_{V_\mathrm{rot}+}) - \log_{10}(V_\mathrm{rot} - \sigma_{V_\mathrm{rot}-}) \:\right) \nonumber \\
& = & \frac{1}{2} \: \left(\:\log_{10}\left[V_\mathrm{rot} \left(1 + \frac{\sigma_{V_\mathrm{rot}+}}{V_\mathrm{rot}}\right)\right]  - \log_{10}\left[V_\mathrm{rot}\left(1 - \frac{\sigma_{V_\mathrm{rot}-}}{V_\mathrm{rot}}\right)\right] \:\right) \nonumber \\
& = & \frac{1}{2} \: \left(\:\log_{10}\left(1 + \frac{\sigma_{V_\mathrm{rot}+}}{V_\mathrm{rot}}\right)  - \log_{10}\left(1 - \frac{\sigma_{V_\mathrm{rot}-}}{V_\mathrm{rot}}\right) \:\right) \nonumber \\
& = & \frac{1}{2} \: \frac{1}{\ln(10)} \: \left(\: \frac{\sigma_{V_\mathrm{rot}+}}{V_\mathrm{rot}} + \frac{\sigma_{V_\mathrm{rot}-}}{V_\mathrm{rot}} \:\right) \nonumber \\
& = & \frac{1}{\ln(10)} \: \frac{(\sigma_{V_\mathrm{rot}+} + \sigma_{V_\mathrm{rot}-})/2}{V_\mathrm{rot}} \;\;\; .
\label{eqn:log_vel_err_symm_simplified}
\end{eqnarray}

\noindent
In other words, the symmetrized velocity error in logarithmic units is obtained from the arithmetic mean of the lower and upper velocity errors in linear units. At this point we have all quantities of interest and their associated observational errors expressed in logarithmic units (Eqns. \ref{eqn:log_vrot}-\ref{eqn:log_mbar} \& \ref{eqn:log_mbar_err}-\ref{eqn:log_vel_err_symm_simplified}). Errors are assumed to be Gaussian in log-space, or equivalently log-normal in linear space. We then seek the linear model for the BTFR that best describes the observational datapoints.

We parametrize our linear model as

\begin{eqnarray}
m_b = \alpha \times (v - \bar{v}) + \beta \;\;\; .
\label{eqn:linear_model}
\end{eqnarray}

\noindent
In the equation above, $\alpha$ is the slope of the line and beta is the intercept at the ``pivot value'' of velocity, $\bar{v}$. We use as the pivot value the average rotational velocity (in logarithmic units) for the sample of galaxies entering our fit, 

\begin{eqnarray}
\bar{v} = \frac{1}{N} \: \sum_i v_i \;\;\; .
\label{eqn:vbar}
\end{eqnarray}

\noindent
Since our observational datapoints possess errors in both the $x$ and $y$ axes, neither a forward linear fit (i.e., ignoring the velocity errors) nor an inverse fit (i.e., ignoring the baryonic mass errors) is appropriate. Instead we look for the linear fit that best reproduces the perpendicular distances of datapoints from the fit line, $d_{\perp,i}$. 
Accordingly, the relevant error for the fit is $\sigma_{\perp,i}$, obtained by projecting the errors on both mass and velocity onto the direction perpendicular to the fit line. More specifically, if $\delta$ is the angle between the fit line and the $x$-axis then the perpendicular distance and perpendicular error are given by:

\begin{eqnarray}
d_{\perp,i} & = & \left[ m_{b,i} - \alpha (v_i - \bar{v}) - \beta \right] \cdot \cos\delta \;\;\; \mathrm{and} \label{eqn:dist_perp} \\
\sigma_{\perp,i}^2 & = & \left( \sigma_{m_b,i} \cdot \cos\delta \right)^2 + \left( \sigma_{v,i} \cdot \sin\delta \right)^2  \;\;\; . 
\label{eqn:sigma_perp}
\end{eqnarray}

\noindent
Note that the angle $\delta$ is related to the linear slope $\alpha$ through the simple relation $\alpha = \tan\delta$. As a result, Eqns. \ref{eqn:dist_perp} \& \ref{eqn:sigma_perp} can be rewritten as:

\begin{eqnarray}
d_{\perp,i} & = & \left[ m_{b,i} - \alpha (v_i - \bar{v}) - \beta \right] \times \frac{1}{\sqrt{1+\alpha^2}} \;\;\;\; \mathrm{and} \label{eqn:dist_perp_simpl} \\
\sigma_{\perp,i}^2 & = & \left( \sigma_{m_b,i} \cdot \frac{1}{\sqrt{1+\alpha^2}} \right)^2 + \left( \sigma_{v,i} \cdot \frac{\alpha}{\sqrt{1+\alpha^2}} \right)^2  \;\;\; .
\label{eqn:sigma_perp_simpl}
\end{eqnarray}


Now we can go ahead and calculate the likelihood of observing datapoint $i$ at a perpendicular distance $d_{\perp,i}$ given the model of Eqn. \ref{eqn:linear_model}:

\begin{eqnarray}
\ell_i \,(\alpha,\beta) = \frac{1}{\sigma_{\perp,i}\sqrt{2\pi}} \times \exp{\left( -\frac{d_{\perp,i}^2}{2\sigma_{\perp,i}^2} \right)} \;\;\; .
\label{eqn:individual_likelihood}
\end{eqnarray}

\noindent
Consequently, the joint likelihood of observing all datapoints in the sample is 

\begin{eqnarray}
\mathcal{L}\, (\alpha,\beta) = \prod_i \ell_i = \prod_i \: \frac{1}{\sigma_{\perp,i}\sqrt{2\pi}} \cdot \exp{\left( -\frac{d_{\perp,i}^2}{2\sigma_{\perp,i}^2}\right)} \;\;\; .
\label{eqn:joint_likelihood}
\end{eqnarray}

\noindent
Our goal is to find the parameters $\alpha$ and $\beta$ that maximize the joint likelihood above. The terms $1/(\sigma_{\perp,i}\sqrt{2\pi})$ are dependent on the model parameters (namely on $\alpha$), and therefore have to be included in the process of likelihood maximization. As a result, likelihood maximization in the case of an orthogonal fit is not equivalent to ``$\chi^2$ minimization'' of the exponents in Eqn. \ref{eqn:joint_likelihood} (in contrast to the case of fits along the vertical or horizontal directions).

Let us also point out that the form of Eqns. \ref{eqn:joint_likelihood} and \ref{eqn:sigma_perp} reflects the assumptions made here that errors among different galaxies are uncorrelated, and that errors in the velocity and baryonic mass of each galaxy are also uncorrelated. These assumptions are simplifying approximations, since in reality neither of them is expected to hold. For example, systematic errors in stellar mass estimates or the absolute calibration of HI fluxes are likely to be shared by all galaxies. In addition, errors on \mbar \ and \vrot \ are expected to be correlated to some degree due to the gas-dominated nature of \gdtf \ galaxies. More specifically, baryonic masses for \gdtf \ galaxies are determined mostly by the measurement of HI flux from their ALFALFA spectra, which are the same spectra used to derive \vrot \ from the width of the HI line profile.

\begin{figure}[htb]
\centering
\includegraphics[width=\linewidth]{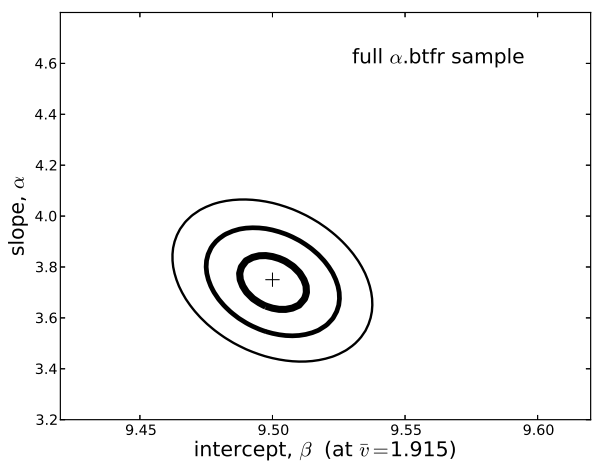}
\caption{ 
Likelihood contours in the slope-intercept plane, for the linear fit to the full \gdtf \ sample.
The cross denotes the maximum likelihood pair of values for the slope, $\alpha$, and intercept, $\beta$ (Eqn. \ref{eqn:linear_model}; refer also to Fig. \ref{fig:btfr}). The solid lines denote the $1\sigma$, $2\sigma$ and $3\sigma$ error regions for the parameters in the $\{\alpha,\beta\}$ plane. 
}
\label{fig:chi2_ellipse}
\end{figure}

\begin{figure}[htb]
\centering
\includegraphics[scale=0.50]{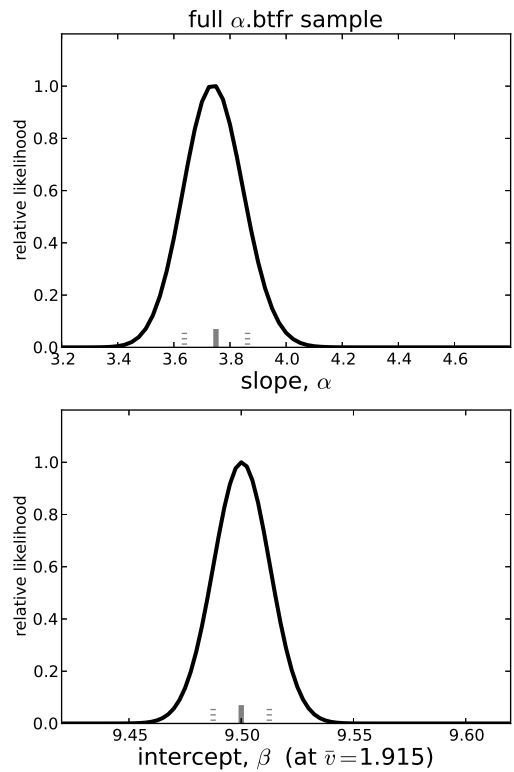}
\caption{ 
\textit{top panel}: Marginalized likelihood distribution for the slope parameter, $\alpha$, of the linear fit to the full \gdtf \ sample. The vertical solid and dotted marks denote the maximum likelihood value and $1\sigma$ range, $\alpha = 3.74 \pm 0.11$. \textit{bottom panel}: Same as the top panel but for the intercept parameter, $\beta$. The maximum likelihood value and $1\sigma$ range for the parameter is $\beta = 9.500 \pm 0.013$, and refers to the pivot value $\bar{v} = 1.915$. 
}
\label{fig:parameter_likelihood}
\end{figure}

\begin{figure}[htb]
\centering
\includegraphics[width=\linewidth]{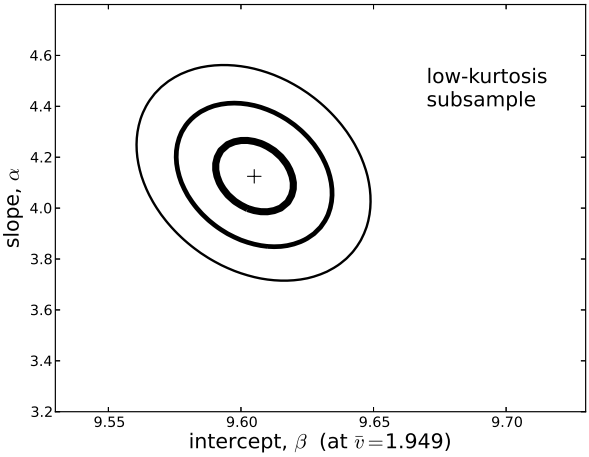}
\caption{ 
Same as Figure \ref{fig:chi2_ellipse}, but for the case of the low-kurtosis subsample of \gdtf \ galaxies (refer to Fig. \ref{fig:btfr_kurt}). Note that the intercept parameter, $\beta$, refers to a pivot value of $\bar{v} = 1.949$, which is different from the pivot value for the full \gdtf \ sample.
}
\label{fig:chi2_ellipse_dh}
\end{figure}

\begin{figure}[H]
\centering
\includegraphics[scale=0.50]{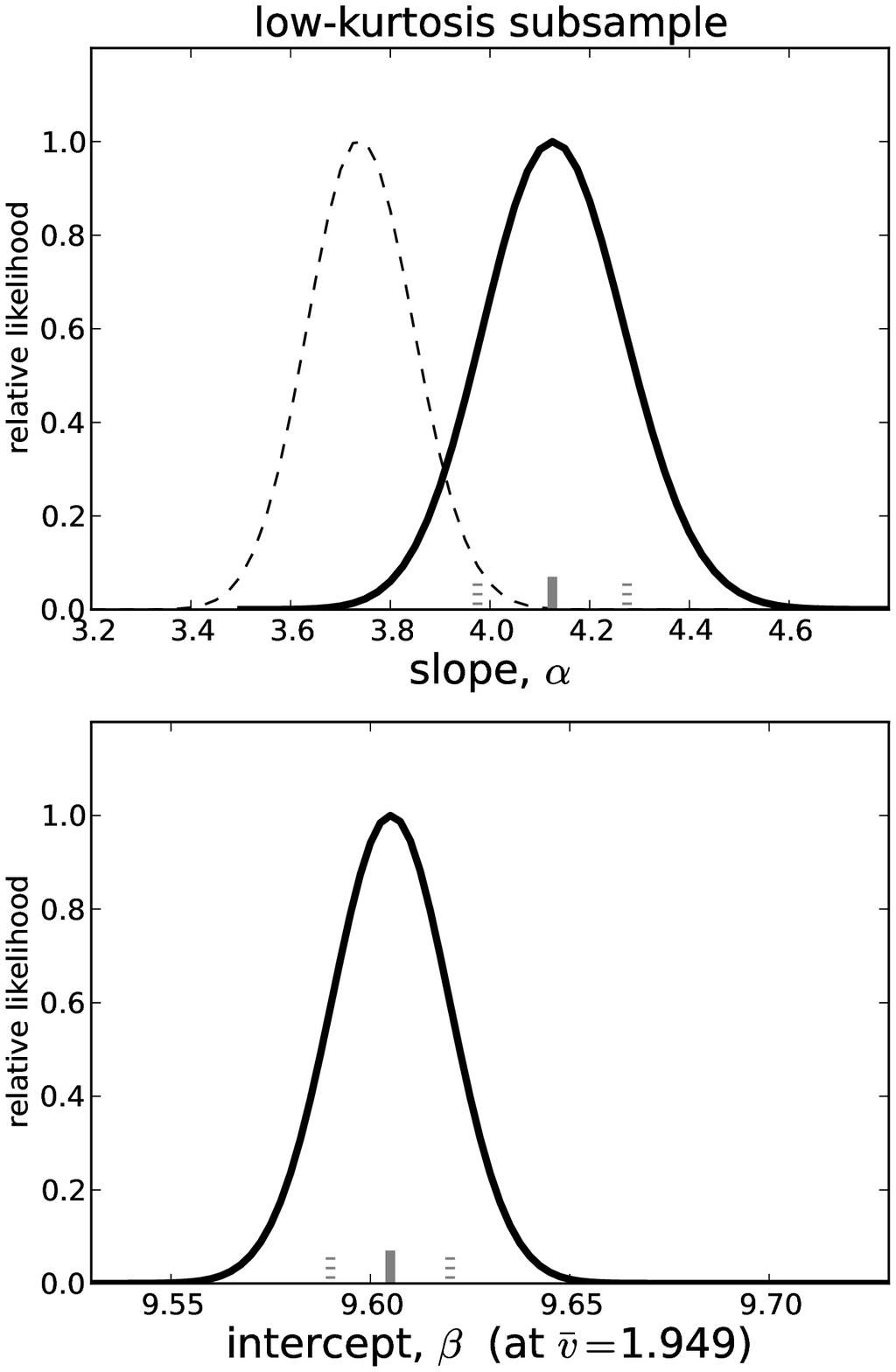}
\caption{ 
\textit{top panel}: Same as the top panel of Figure \ref{fig:parameter_likelihood}, but for the case of the low-kurtosis subsample of \gdtf \ galaxies. The maximum likelihood slope for this subsample is $\alpha = 4.14 \pm 0.14$. The likelihood distribution of the slope for the full \gdtf \ sample is also plotted as a thin dashed line, for comparison. \textit{bottom panel}: Same as the bottom panel of Figure \ref{fig:parameter_likelihood}, but for the case of the low-kurtosis subsample of \gdtf \ galaxies. The maximum likelihood intercept for this subsample is $\beta = 9.605 \pm 0.015$, and refers to a pivot value of $\bar{v} = 1.949$. Keep in mind that the pivot values for the full \gdtf \ sample and for the low-kurtosis subsample are different.   
}
\label{fig:parameter_likelihood_dh}
\end{figure}

Figure \ref{fig:chi2_ellipse} shows the results of the fitting procedure described above for the full \gdtf \ sample (refer to Fig. \ref{fig:btfr}). Keep in mind that the three outlier galaxies identified in Figure \ref{fig:btfr} are excluded from the fit. The figure shows the maximum likelihood values of the parameters $\alpha$ and $\beta$ of the linear fit to the BTFR, together with their $1\sigma$, $2\sigma$ and $3\sigma$ contours in the $\{\alpha,\beta\}$ plane. Figure \ref{fig:parameter_likelihood} shows instead the marginalized one-dimensional likelihood distributions separately for the two parameters. From these marginalized distributions we obtain the best fit parameter values and their errors, as reported in Sec. \ref{sec:btfr}: $\alpha = 3.74 \pm 0.11$ and $\beta = 9.500 \pm 0.013$ (referring to the pivot value $\bar{v} = 1.915$). Figures \ref{fig:chi2_ellipse_dh} and \ref{fig:parameter_likelihood_dh} are analogous to Figs. \ref{fig:chi2_ellipse} and \ref{fig:parameter_likelihood}, but they now refer to the low-kurtosis subsample of \gdtf \ galaxies (refer to Fig. \ref{fig:btfr_kurt}). The best fit slope for the low-kurtosis subsample is $\alpha = 4.14 \pm 0.14$, slightly steeper than the value obtained by fitting the full sample. The best fit intercept for the low-kurtosis subsample is $\beta = 9.605 \pm 0.015$, and refers to the pivot value $\bar{v} = 1.949$. Keep in mind that the slope parameter values obtained for the full \gdtf \ sample and the low-kurtosis subsample can be directly compared, but the intercept parameter values not (they refer to different velocity pivot points).

\subsection{Alternative fitting methodologies}
\label{sec:alternative_fits}

One major obstacle in the scientific analysis of the BTFR is that the linear fits to the relation are not always performed according to the same methodology. This creates a lot of confusion, since different fitting methods can result in different best fit parameter values, even in the case that the same observational data are used (see, e.g., detailed analysis by \citealp{Bradford2016}).


\begin{table*}[htb]
    \centering
    \begin{tabular}{llccc}
        
        \hline
        
        fit method & code & slope, $\alpha$ & intercept, $\beta$ & intrinsic scatter, $\sigma_{\perp,\mathrm{intr}}$ [dex] \\
        
         & & & (at pivot velocity $\bar{v}$) & (in perpend. direction) \\ [0.5ex]
        
        \hline \hline \\
        
        \multicolumn{5}{c}{\textbf{full \gdtf \ sample}  ($\bar{v} = 1.915$)}\\ [0.5ex]
        
        \hline

        orthogonal ML (no scatter)$^\ast$ &  this work  & 3.75 $\pm$ 0.11  & 9.500 $\pm$ 0.013 & 0  \\
       
        orthogonal ML (scatter free) &  this work &  4.15 $\pm$ 0.23  & 9.493 $\pm$ 0.025 & 0.045 $\pm$ 0.005 \\

        forward &  \texttt{bces} &  3.50 $\pm$ 0.18  & 9.498 $\pm$ 0.022 & ... \\
 
        inverse &  \texttt{bces} &  4.01 $\pm$ 0.23  & 9.498 $\pm$ 0.024 & ... \\

        bisector &  \texttt{bces} &  3.74 $\pm$ 0.18  & 9.498 $\pm$ 0.023 & ... \\

        orthogonal &  \texttt{bces} &  3.98 $\pm$ 0.22  & 9.498 $\pm$ 0.023 & ... \\
        
        \hline \\
        
        \multicolumn{5}{c}{\textbf{low-kurtosis subsample}  ($\bar{v} = 1.949$)} \\
        [0.5ex]
        
        \hline
        
        orthogonal ML (no scatter)$^\ast$ &  this work  & 4.13 $\pm$ 0.15  & 9.605 $\pm$ 0.015 & 0 \\
 
        orthogonal ML (scatter free) &  this work &  4.78 $\pm$ 0.33  & 9.593 $\pm$ 0.030 & 0.041 $\pm$ 0.006 \\

        forward &  \texttt{bces} &  3.98 $\pm$ 0.25  & 9.601 $\pm$ 0.026 & ... \\
 
        inverse &  \texttt{bces} &  4.62 $\pm$ 0.29  & 9.601 $\pm$ 0.028 & ... \\

        bisector &  \texttt{bces} &  4.27 $\pm$ 0.23  & 9.601 $\pm$ 0.027 & ... \\

        orthogonal &  \texttt{bces} &  4.58 $\pm$ 0.28  & 9.601 $\pm$ 0.028 & ... \\
        
        \hline

    \end{tabular}
    \caption{Summary of linear fit results for all methods discussed in Appendix \ref{sec:fit_details} \& \ref{sec:alternative_fits}. An asterisk denotes the ``fiducial'' fitting method used to derive the parameter values reported in the main body of this article. The triple-dot denotes parameters whose best fit value and error are not reported by the fitting code.}
    \label{tab:fits}
\end{table*}

The linear fit described in the preceding section is a ``no-scatter fit'', in the sense that intrinsic scatter is not included as a parameter in the linear model of Eqn. \ref
{eqn:linear_model}. We consider this no-scatter fit as our ``fiducial'' fit, and report the resulting best fit parameters in the main body of this article. The reason for adopting this particular fitting methodology as our fiducial method is that no-scatter fits are by far the most widely used BTFR fits in the literature, and this helps us to compare our results to the results of previous studies. In this section, we consider instead a range of alternative fitting methods and summarize the results in Table \ref{tab:fits}.

First, we consider an extension of the fiducial fit, whereby intrinsic scatter is added to the linear model of Eqn. \ref{eqn:linear_model}. As the tests of Figs. \ref{fig:histogram} \& \ref{fig:histogram_dh} have revealed the presence of intrinsic scatter in the BTFR of our galactic sample, this model is more appropriate.
In particular, we include intrinsic scatter as a free fitting parameter to be constrained by the data themselves. The scatter is assumed to be Gaussian in the direction perpendicular to the fit line, such that the linear model is now fully characterized by three parameters:

   \[
     \begin{array}{lp{0.8\linewidth}}
         \alpha  & slope     \\
         \beta               &  intercept at the pivot velocity $\bar{v}$                   \\
        \sigma_{\perp,\mathrm{intr}}         &  intrinsic scatter in the direction perpendicular to the fit line.     
      \end{array}
   \]



The likelihood of observing data point $i$ is now different from Eqn. \ref{eqn:individual_likelihood}, and given by the convolution of two Gaussian distributions, one representing the observational error of the datapoint in question and one representing the intrinsic scatter of the linear model,

\begin{eqnarray}
\ell_i \,(\alpha,\beta,\sigma_{\perp,\mathrm{intr}}) & = &  \nonumber \\ & \int_{d^\prime} & \frac{1}{\sigma_{\perp,i}\sqrt{2\pi}} \cdot \exp{\left( -\frac{(d^\prime - d_{\perp,i})^2}{2\sigma_{\perp,i}^2} \right)} \nonumber \\ 
& \times & \frac{1}{\sigma_{\perp,\mathrm{intr}}\sqrt{2\pi}} \cdot \exp{\left( -\frac{d^{\prime 2}}{2\sigma_{\perp,\mathrm{intr}}^2} \right)} \; \mathrm{d} d^\prime  \;\; .
\label{eqn:individual_likelihood_scatter}
\end{eqnarray}

\noindent
After some tedious algebra, the convolution above reduces to the simple expression:

\begin{eqnarray}
\ell_i \,(\alpha,\beta,\sigma_{\perp,\mathrm{intr}}) & = & \nonumber \\
\frac{1}{\sqrt{2\pi\left( \sigma_{\perp,i}^2 + \sigma_{\perp,\mathrm{intr}}^2 \right) }} & \cdot & \exp{\left( -\frac{d_{\perp,i}^2}{2 \left( \sigma_{\perp,i}^2 + \sigma_{\perp,\mathrm{intr}}^2 \right)} \right)}   \;\;\; .
\label{eqn:individual_likelihood_scatter_simple}
\end{eqnarray}

\noindent
Note that if the intrinsic scatter parameter is forced to be $\sigma_{\perp,\mathrm{intr}} = 0$, then the likelihood above reduces to the likelihood of our fiducial no-scatter fit (Eqn. \ref{eqn:individual_likelihood}).

The next step is to find the set of parameters $\alpha$, $\beta$, and $\sigma_{\perp,\mathrm{intr}}$ that maximize the joint likelihood 

\begin{eqnarray}
\mathcal{L}\, (\alpha,\beta,\sigma_{\perp,\mathrm{intr}}) = \prod_i \ell_i(\alpha,\beta,\sigma_{\perp,\mathrm{intr}}) \;\;\; .
\label{eqn:joint_likelihood_scatter}
\end{eqnarray}

\begin{figure*}[htb]
\centering
\includegraphics[scale=0.48]{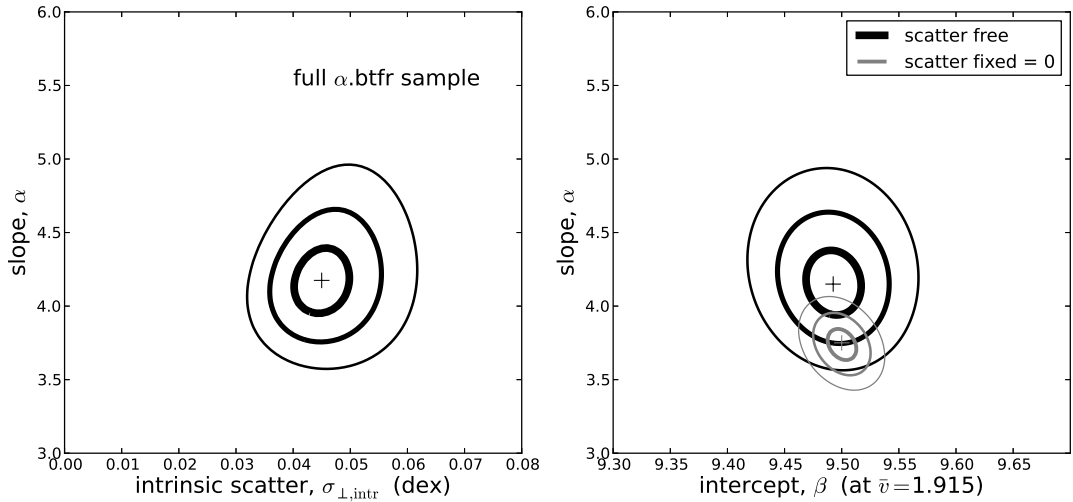}
\caption{ 
Likelihood contours in the slope-intercept-scatter planes, for a linear fit to the full \gdtf \ sample including intrinsic scatter as a free parameter.
\textit{left panel}: The black cross denotes the best fit values of slope, $\alpha$, and intrinsic scatter, $\sigma_{\perp,\mathrm{intr}}$. The solid lines represent the $1\sigma$, $2\sigma$, and $3\sigma$ error regions in the $\alpha$--$\sigma_{\perp,\mathrm{intr}}$ plane. \textit{right panel}: The black cross and black contours represent the best fit value and error regions for the slope, $\alpha$, and intercept, $\beta$. The gray cross and contours represent the results of our ``fiducial'' no-scatter fit to the full \gdtf \ sample (same as Fig. \ref{fig:chi2_ellipse}).  
}
\label{fig:chi2_ellipse_scatter}
\end{figure*}

\noindent
Figure \ref{fig:chi2_ellipse_scatter} shows the result of the likelihood maximization process described above for the full \gdtf \ sample. As the left panel of the figure shows, intrinsic scatter is required by the data to be non-zero at high significance. The best fit value of intrinsic scatter for the full \gdtf \ sample is $\sigma_{\perp,\mathrm{intr}} = 0.045 \pm 0.005$ dex. The right panel of Fig. \ref{fig:chi2_ellipse_scatter} shows instead that introducing scatter to the linear fit affects also the derived best fit values of slope and intercept. More specifically, the best fit parameters for the full \gdtf \ sample are now $\alpha = 4.16 \pm 0.22$ and $\beta = 9.492 \pm 0.025$ (compared with $\alpha = 3.74 \pm 0.11$ and $\beta = 9.500 \pm 0.013$ for the no-scatter fit). In other words, the best fit slope has become noticeably steeper, while the errors on both the slope and intercept have become significantly larger.

\begin{figure}[htb]
\centering
\includegraphics[scale=0.48]{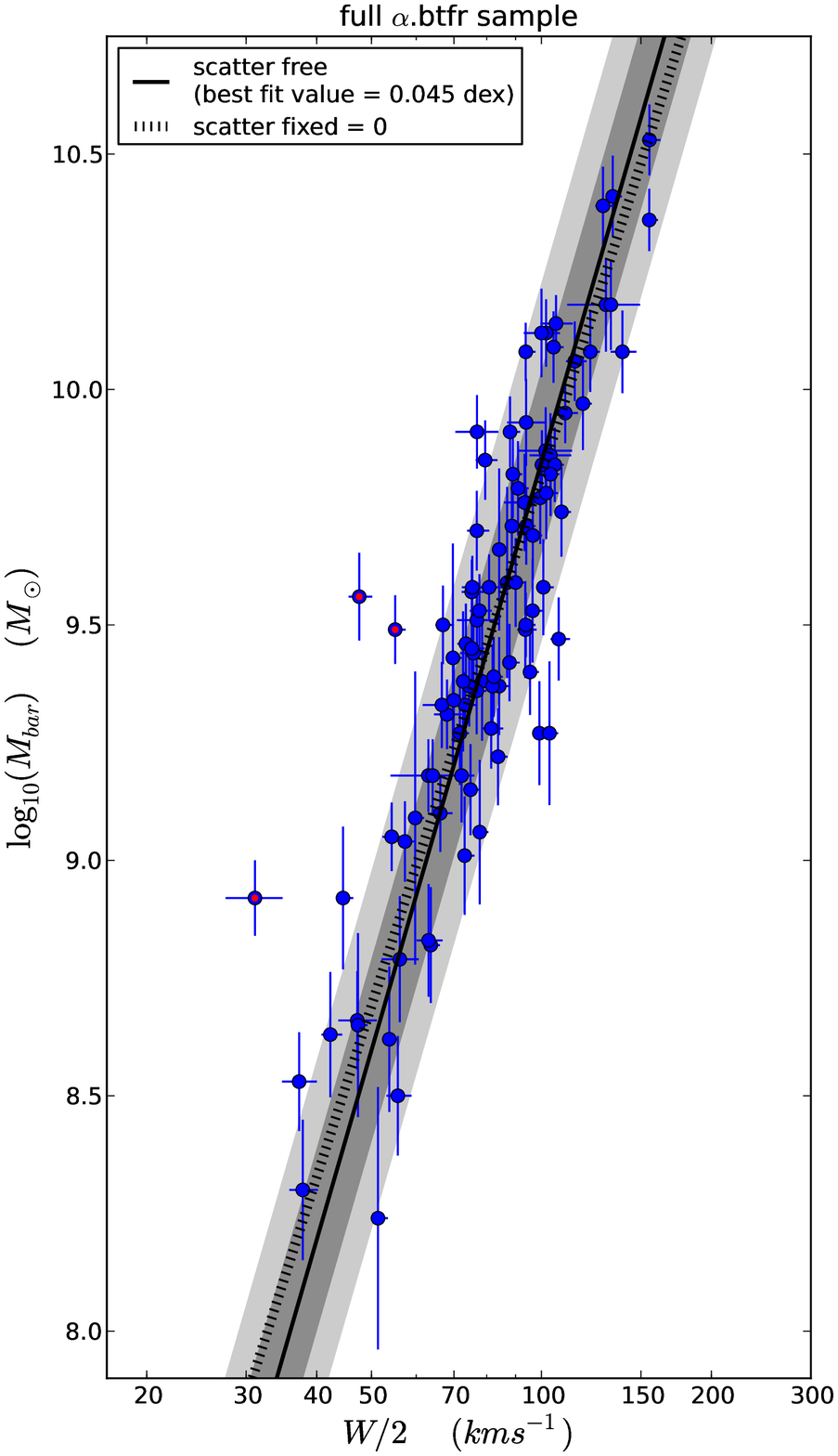}
\caption{ 
Comparison of the linear fit to the BTFR of the full \gdtf \ sample with and without intrinsic scatter included in the model.
Comparison of the linear fit to the BTFR of the full \gdtf \ sample with and without intrinsic scatter included in the model. The layout of the figure follows that of Fig. \ref{fig:btfr}. The thick dotted line represents our fiducial no-scatter fit, and is the same as the thick dotted line in Fig. \ref{fig:btfr}. The thin solid line represents the best fit line when intrinsic scatter is treated as a free parameter. The gray shaded bands represent the 1$\sigma$ and $2\sigma$ intrinsic scatter regions according to the best fit value of the parameter $\sigma_{\perp,\mathrm{intr}} = 0.045$ dex.  
}
\label{fig:btfr_scatter_vs_fixed}
\end{figure}

The effects of introducing intrinsic scatter as a free fitting parameter on the derived values of slope and intercept can be intuitively understood in the following way: In the case of no-scatter, the fit line cannot ``wander'' too much away from datapoints with low observational errors. This is because the contribution to the likelihood from such an object (Eqn. \ref{eqn:individual_likelihood}) drops very quickly as the perpendicular distance between the datapoint and the fit line increases. This translates in the fact that the fit parameters are strongly constrained to be close to their best fit values. In the case of a fit that includes intrinsic scatter, the ``effective'' variance in the individual likelihood terms can be significantly larger than the observational error (see Eqn. \ref{eqn:individual_likelihood_scatter_simple}). Therefore the likelihood contribution of an object with small observational errors drops much more slowly as the distance between the fit line and the datapoint increases, resulting in a larger error range for the fit parameters. The steepening of the slope can also be understood based on the consideration above, once the position of \gdtf \ galaxies on the BTFR diagram is taken into account (see Figure \ref{fig:btfr_scatter_vs_fixed}). The likelihood of the no-scatter fit gets ``rewarded'' more for accommodating the position of some high-mass \gdtf \ galaxies with low observational errors. Given the position of these objects relative to the rest of the \gdtf \ datapoints, the no-scatter fit produces a slightly shallower slope than the fit including intrinsic scatter. In the latter case, the position of these galaxies is interpreted as due to the relation's intrinsic scatter, and the ``reward'' to reproduce their position is greatly reduced.

\begin{figure*}[htb]
\centering
\includegraphics[scale=0.48]{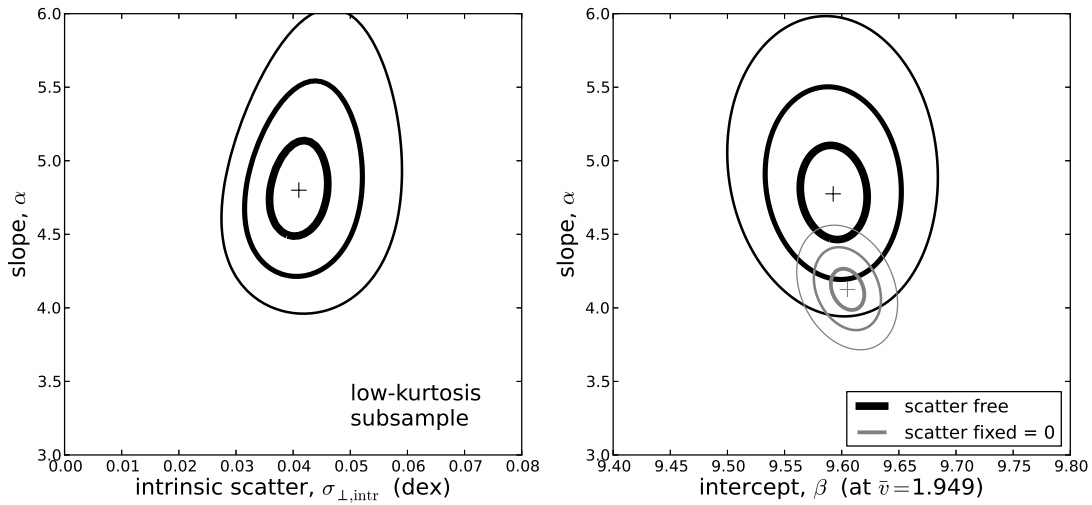}
\caption{ 
Same as Fig. \ref{fig:chi2_ellipse_scatter}, but now referring to the low-kurtosis subsample of \gdtf \ galaxies.   
}
\label{fig:chi2_ellipse_scatter_dh}
\end{figure*}

\begin{figure}[htb]
\centering
\includegraphics[scale=0.48]{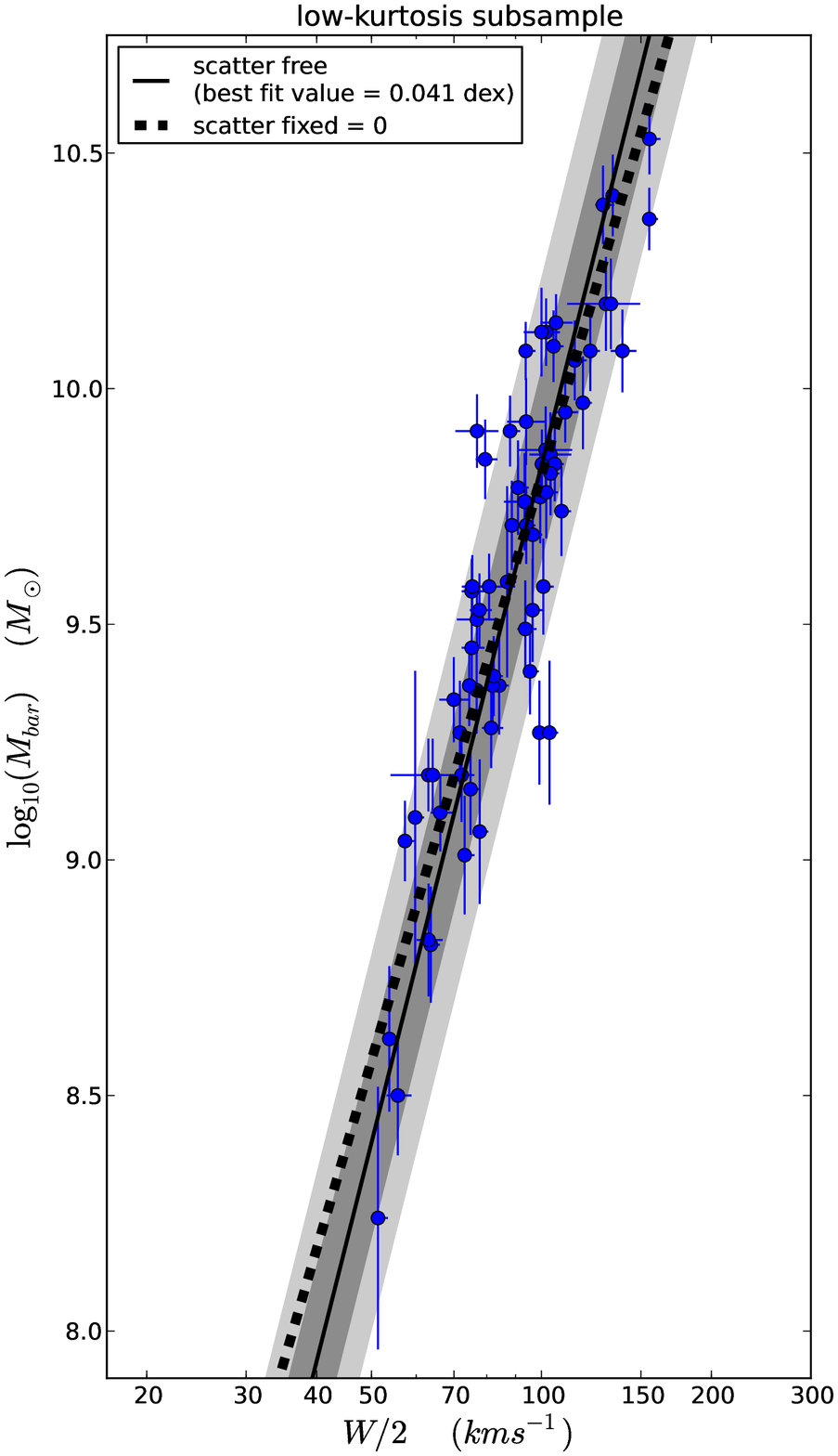}
\caption{ 
Same as Fig. \ref{fig:btfr_scatter_vs_fixed}, but now referring to the low-kurtosis subsample of \gdtf \ galaxies. The thick dashed line is the no-scatter fit to the subsample, and is the same as the thick dashed line in Fig. \ref{fig:btfr_kurt}. The intrinsic scatter bands correspond to the best fit value of the parameter for the subsample, $\sigma_{\perp,\mathrm{intr}} = 0.041$ dex.   
}
\label{fig:btfr_scatter_vs_fixed_dh}
\end{figure}

Figures \ref{fig:chi2_ellipse_scatter_dh} and \ref{fig:btfr_scatter_vs_fixed_dh} show the results of the linear fit including scatter to the subsample of \gdtf \ galaxies with low profile kurtosis (refer to \S\ref{sec:profile_shape}). Qualitatively, the trends observed are very similar to those seen for the full \gdtf \ sample (Figs. \ref{fig:chi2_ellipse_scatter} \& \ref{fig:btfr_scatter_vs_fixed}). Intrinsic scatter is detected once again at high significance, and the resulting best fit value is $\sigma_{\perp,\mathrm{intr}} = 0.041 \pm 0.005$ dex. The best fit slope and intercept are now $\alpha = 4.78 \pm 0.33$ and $\beta = 9.595 \pm 0.030$ (compared to $\alpha = 4.14 \pm 0.14$ and $\beta = 9.605 \pm 0.015$ for the no-scatter fit). Note that while the slope of the no-scatter fit is compatible within 1$\sigma$ with the slope predicted by MOND ($\alpha = 4$), the slope of the fit including scatter as a free parameter is incompatible at the $\approx 2.5\sigma$ level.


Lastly, we consider fits to the \gdtf \ sample performed with the \texttt{bces} code, written in the Python programming language\footnotemark{}. The code is based on the statistical model developed in \citet{AkritasBershady1996}, and can be applied to datasets with (possibly correlated) errors in both the dependent and independent variables. The parameters of the best fit line and their errors are estimated by \texttt{bces} based on bootstrapping the observational dataset. The code calculates linear fits of four different types: forward ($M_\mathrm{bar}|V_\mathrm{rot}$), inverse ($V_\mathrm{rot}|M_\mathrm{bar}$), bisector, and orthogonal. Here we report the \texttt{bces} orthogonal fit results, since this type of fit is the closest to our orthogonal ML fits described in Appendix \ref{sec:fit_details} \& \ref{sec:alternative_fits}. For the full \gdtf \ sample \texttt{bces} measures a slope of $\alpha = 3.98 \pm 0.22$, while for the low-kurtosis subsample, $\alpha^{(\mathrm{low-}k)} = 4.58 \pm 0.28$. We can see that the \texttt{bces} results are more similar to the results obtained in this work when intrinsic scatter is included in the fit, rather than to our ``fiducial'' no-scatter linear fit.

\footnotetext{\texttt{https://github.com/rsnemmen/BCES}}

\end{document}